\documentclass[journal,a4paper,twoside]{IEEEtran}

\IEEEoverridecommandlockouts

\newcommand\hmmax{0}
\newcommand\bmmax{0}

\usepackage{graphicx}
\usepackage{tikz}
\usepackage{pgfplots}
\usepackage{xcolor}
\usepackage{color, colortbl}
\usepackage{amsmath}
\usepackage{bbm}
\usepackage{amsfonts, amssymb}
\usepackage{mathtools}
\usepackage{cite}
\usepackage[nolist]{acronym}

\usepackage{booktabs} 		
\usepackage{diagbox}		
\usepackage{upgreek}
\usepackage{bbm}
\usepackage{Files/bm}
\usepackage{Files/winsnotation}
\usepackage{Files/Symbols_OPL}
\usepackage{Files/LVcolours} 

\usepackage[linesnumbered,ruled,vlined]{algorithm2e}
\usepackage{setspace}	 	
\colorlet{AlgCaptionColor}{gray!5}
\makeatletter
\renewcommand{\algocf@makecaption@ruled}[2]{%
  \global\sbox\algocf@capbox{\colorbox{AlgCaptionColor}{\hskip\AlCapHSkip
    \parbox[t]{\hsize}{\algocf@captiontext{\strut#1}{\strut#2\strut}}\hskip1.5\algomargin}}
}%
\makeatother
\definecolor{ferngreen}{rgb}{0.31, 0.47, 0.26}

\SetCommentSty{mycommfont}
\usepackage{comment}

\usepackage{array}
\newcolumntype{C}[1]{>{\centering\arraybackslash}p{#1}}

\makeatletter
\newcommand\notsotiny{\@setfontsize\notsotiny\@viipt\@viiipt}
\makeatother

\makeatletter
\def\endthebibliography{%
  \def\@noitemerr{\@latex@warning{Empty `thebibliography' environment}}%
  \endlist
}
\makeatother
\IEEEoverridecommandlockouts
\usepackage{amsthm}
\theoremstyle{definition}
\newtheorem{remark}{Remark}
\newtheorem{example}{Example}

\pgfplotsset{compat=1.17}

\begin{document}

\title{Interference Cancellation Algorithms\\ for Grant-Free Multiple Access \\ with Massive MIMO}

\author{Lorenzo~Valentini,~\IEEEmembership{Graduate~Student~Member,~IEEE,}
        Marco~Chiani,~\IEEEmembership{Fellow,~IEEE,}
        and~Enrico~Paolini,~\IEEEmembership{Senior~Member,~IEEE}

\thanks{The authors are with the Department of Electrical, Electronic, and Information Engineering ``Guglielmo Marconi'' and CNIT/WiLab, University of Bologna, 40136 Bologna, Italy. E-mail: \{lorenzo.valentini13, marco.chiani, e.paolini\}@unibo.it.} 
\thanks{This work has been presented in part at IEEE International Conference on Communications, Seoul, South Korea, May 2022.}
}

\maketitle 

\begin{acronym}
\small
\acro{ACK}{acknowledgement}
\acro{AWGN}{additive white Gaussian noise}
\acro{BCH}{Bose–Chaudhuri–Hocquenghem}
\acro{BS}{base station}
\acro{CDF}{cumulative distribution function}
\acro{CHB}{channel-hardening-based}
\acro{CRA}{coded random access}
\acro{CRC}{cyclic redundancy check}
\acro{CRDSA}{contention resolution diversity slotted ALOHA}
\acro{CSA}{coded slotted ALOHA}
\acro{eMBB}{enhanced mobile broad-band}
\acro{FER}{frame error rate}
\acro{IFSC}{intra-frame spatial coupling}
\acro{i.i.d.}{independent and identically distributed}
\acro{IoT}{Internet-of-Things}
\acro{IRSA}{irregular repetition slotted ALOHA}
\acro{LDPC}{low-density parity-check}
\acro{LOS}{line of sight}
\acro{MAC}{medium access control}
\acro{MIMO}{multiple input multiple output}
\acro{ML}{maximum likelihood}
\acro{MMA}{massive multiple access}
\acro{mMTC}{\emph{massive} machine-type communication}
\acro{MTC}{machine-type communication}
\acro{MPR}{multi-packet reception}
\acro{MRC}{maximal ratio combining}
\acro{PAB}{payload-aided-based}
\acro{PDF}{probability density function}
\acro{PHY}{physical}
\acro{PLR}{packet loss rate}
\acro{PMF}{probability mass function}
\acro{PRCE}{perfect replica channel estimation}
\acro{QAM}{quadrature amplitude modulation}
\acro{QPSK}{quadrature phase-shift keying}
\acro{RF}{radio-frequency}
\acro{SC}{spatial coupling}
\acro{SIC}{successive interference cancellation}
\acro{SIS}{successive interference subtraction}
\acro{SNR}{signal-to-noise ratio}
\acro{URLLC}{ultra-reliable and low-latency communication}
\end{acronym}
\setcounter{page}{1}

\begin{abstract}
In next generation Internet-of-Things, the overhead introduced by grant-based multiple access protocols may engulf the access network as a consequence of the unprecedented number of connected devices. 
Grant-free access protocols are therefore gaining an increasing interest to support massive access from machine-type devices with intermittent activity. 
In this paper, coded random access (CRA) with massive multiple input multiple output (MIMO) is investigated as a solution to design highly-scalable massive multiple access protocols, taking into account stringent requirements on latency and reliability.
With a focus on signal processing aspects at the physical layer and their impact on the overall system performance, critical issues of successive interference cancellation (SIC) over fading channels are first analyzed. Then, SIC algorithms and a scheduler are proposed that can overcome some of the limitations of the current access protocols. 
The effectiveness of the proposed processing algorithms is validated by Monte Carlo simulation, for different CRA protocols
and by comparisons with developed benchmarks.
\end{abstract}

\begin{keywords}
Coded random access, grant-free access, massive MIMO, massive multiple access, signal processing, successive interference cancellation.
\end{keywords}

\section{Introduction}


The rise of the \ac{IoT} has progressively furthered attention on \ac{MTC}, meant as the autonomous communication between physical objects not directly operated by humans \cite{Sachs2016:Machine}. 
Due to the fast growing \ac{IoT} pervasiveness in several application domains, the density of connected objects (in terms of devices per unit area) has recently become so large that the expression \ac{mMTC} has been introduced \cite{Shariatmadari2015:Machine,Bockelmann2016:Massive} to indicate wireless networking among a very large number of devices that are physically located in the same area.
Each of these devices, usually battery-driven, generates data discontinuously and intermittently; the device's activity periods are typically used for transmission of one message, in the form of a short packet, to another device or to a remote server through the network.
A usual situation is the one where a massive number of devices are connected wirelessly to the network through the same \ac{BS}. 
As such, in the uplink a typical \ac{MMA} problem occurs, in which a myriad of transmitters contend to transmit short data packets to the same receiver over the radio access network \cite{Lien2011:Toward,Wu2020:Massive}. 
Since transmitters wake up intermittently, unpredictably, and independently of each other, the receiver has no a priori knowledge of the number and the subset of simultaneously active ones within a given time interval.

It should be remarked that the \ac{MMA} setting deviates significantly from the traditional multiple access one.
As opposed to the conventional context, where orthogonal channel access is feasible owing to the relatively small number of transmitters and where grant-based protocols with pre-allocation of radio resources are justified by each transmitter typically holding the assigned resources for a while, in \ac{MMA} applications scheduled access schemes are very inconvenient and inefficient, with control signaling that may even outnumber data. 
Moreover, from a theoretical point of view, the traditional information-theoretic tools to analyze multi-access communication are insufficient to address fundamental limits of massive access.
This problem, that has been known for a long time \cite{wolf1981:coding,Gallager1985:Perspective,Mathys1990:class},
has recently received a renewed interest \cite{Durisi2016:Toward,Polyanskiy2017:Perspective,Chen2017:Capacity,Ngo2021:Random_user_activity,Paolini22:Irregular}.

The main challenge for next-generation \ac{MMA} schemes is represented by the need of featuring a very high scalability, in terms of capability to support the ever-increasing connection densities, in presence of reliability and latency constraints that, although smoother than those characterizing \ac{URLLC} services, may be much more tightening than the typical \ac{mMTC} 5G ones (packet loss probability not exceeding $1$\% and latency not exceeding $10\,$s) \cite{Hasan2013:random,Liu2018:sparse,Chen2020:massive,Gui2020:6G,Kalalas2020:massive,Pokhrel2020:Towards}. 
To cope with these requirements, next-generation \ac{MMA} schemes should be designed to maximize the number of \emph{simultaneously active} machine-type devices (or ``users''), each contending for transmission of one data packet, for which a target reliability and a target latency can be guaranteed with no a priori knowledge of the users' activation pattern. 
In this respect, grant-free multiple access schemes, with no pre-existing resource allocation or handshake procedure between the user and the \ac{BS}, have recently gained an increasing interest owing to their capability to substantially reduce control signalling for connection establishment, which is beneficial in terms of scalability and latency, as well as of protocol lightness and energy efficiency on the device side.
Grant-free access schemes, on the other hand, tend to increase complexity at the \ac{PHY} layer on the receiver side, due to the need to perform packet detection and also channel estimation directly from the detected packets.
Examples of grant-free schemes are the ones recently proposed in \cite{Liu2018:massivePt1,Sor2018:coded,Fengler2019:grant-free,Han2020:grant-free,Abebe2021:MIMO,Choi2022:grant,Decurninge2021:Tensor}\cite{Liu22:unsourced}.
Typical \ac{MMA} schemes are also uncoordinated, meaning that simultaneously active devices take actions independently of each other, without any coordination or cooperation.

Uncoordinated protocols based on the \ac{CRA} paradigm \cite{casini2007:contention,liva2011:irsa,paolini2015:csa,paolini2015:magazine,clazzer2018:combining,Berioli2016:Modern,Munari2021:age,ValChiPao:22}, a particular class of grant-free access schemes, ensure high reliability and are currently regarded as candidates for 6G \cite{Mahmood2020:Six} due to their capability of bridging random access with iterative decoding of codes on sparse graphs.
Some of these protocol, e.g., \acl{CRDSA} \cite{casini2007:contention} or \acl{IRSA} \cite{liva2011:irsa}, are based on packet repetition; some others, like \ac{CSA} \cite{paolini2015:csa}, on packet fragmentation and packet-level coding. 
Packet replicas or coded fragments are transmitted on different resources, and resource diversity is combined with interference cancellation performed by simple \acl{SIS} at the receiver \cite{paolini2015:csa}.
As a matter of fact, the performance of \ac{CRA} schemes does not depend only on the \ac{MAC} protocol, e.g., the uncoordinated resource selection strategy on the device side; it also heavily relies on the effectiveness of the \ac{PHY} layer processing algorithms at the receiver.
Although part of the literature on \ac{CRA} tends to model the \ac{PHY} layer signal processing (including packet detection, channel estimation, and interference cancellation) as ideal, signal processing in a realistic setting may introduce considerable losses with respect to the performance under idealized conditions, especially in terrestrial scenarios characterized by fading. 
For instance, the often employed collision or \ac{MPR} channel models \cite{Gha2013:irregular,stefanovic2018:multipacket} may sometimes turn inaccurate, jeopardizing effective system design and optimization \cite{Valentini2022:Joint}. 

In this paper, we address \ac{SIC} algorithms at \ac{PHY} layer for \ac{CRA} schemes in massive access applications. 
We start by critically reviewing a low-complexity \ac{SIC} algorithm proposed in \cite{Sor2018:coded} and tailored to massive \ac{MIMO} processing at the receiver; an in-depth analysis for this algorithm is developed and possible vulnerabilities are highlighted.
Motivated by this analysis, we then propose an innovative massive \ac{MIMO} \ac{SIC} algorithm that is able to considerably improve the number of supported simultaneously active devices for given reliability and latency constraints.
The algorithm relies on two main observations. 
The first one is that, in \ac{CRA} protocols, it is possible to effectively exploit resource diversity to accurately estimate the channel coefficients in the resources in which interference must be subtracted. 
The second one is that not all interference cancellation operations are equally effective: Introducing a prioritization to schedule the most effective ones first is expected to improve the overall performance.
The key contributions of the paper can be summarized as follows:
\begin{itemize}
    \item we exploit resource diversity and operation scheduling to improve \ac{SIC} in presence of a massive number of \ac{BS} antennas;
    \item we theoretically analyze the interference effects within a slot, providing system design guidelines;
    \item we investigate scalability of several \ac{PHY} and \ac{MAC} protocol configurations.
\end{itemize}

This paper is organized as follows. Section~\ref{sec:preliminary} introduces preliminary concepts, the system model, and some background material. Section~\ref{sec:ImprovingSIC} describes the proposed \ac{SIC} technique along with an analysis 
that justifies the gain introduced by the new scheme. 
Numerical results are shown in Section~\ref{sec:NumericalResults}. Finally, conclusions are drawn in Section~\ref{sec:conclusions}.
A subset of the results presented in this work appeared in the conference paper \cite{Valentini2022:Impact}. 
With respect to \cite{Valentini2022:Impact}: (i) the proposed schemes are addressed in a more thorough way, providing all details and extending the analysis to include noise (besides interference) and to generic $\mathsf{M}$-\ac{QAM} constellations; (ii) the concept of cancellation scheduling is introduced to further improve performance; (iii) performance benchmarks are obtained and used as a reference in the numerical results; (iv) richer numerical results are presented.

\emph{Notation}: Throughout the paper, capital and lowercase bold letters denote matrices and vectors, respectively. The conjugate transposition of a matrix or vector is denoted by $(\cdot)^H$, while $\|\cdot\|$ indicates the Euclidean norm. The operator $\E{\cdot}$ denotes expectation, while $\Var{\cdot}$ is used for variance. 

\section{Preliminaries and Background}
\label{sec:preliminary}
In this section we define the reference scenario, including the channel access protocols 
and the channel model, also reviewing some \ac{PHY} layer signal processing techniques performed at the receiver, that will be useful in the sequel.

\subsection{Scenario Definition}\label{subsec:Scenario}

We consider an scenario with $K$ single-antenna users ($K$ very large) which aim at transmitting simple uplink messages to one receiving \ac{BS} equipped with multiple antennas.
The \ac{BS} time is organized into frames, with a periodic beacon signal broadcast by the \ac{BS} at the beginning of each frame.
The frames are divided into $N_\mathrm{s}$ slots, and users are frame- and slot-synchronous by relying on the beacon signal.
The $K$ users wake up unpredictably to transmit data in a frame and therefore they are not all simultaneously active. 
The number of simultaneously active users, contending for transmission of their packet in the same frame, is denoted by $K_{\mathrm{a}}$ and we assume that the receiver has no prior knowledge about this number. 
The $K_\mathrm{a}$ active users contend for the channel in a grant-free and uncoordinated fashion to send their uplink data to the \ac{BS}.

In this paper, we consider the channel access protocols belonging the class of \ac{CSA} \cite{paolini2015:csa}. 
We focus on the specific case of \ac{CSA} with repetition codes of a given rate $1/r$ for all users. 
This means that each active user generates $r$ replicas of its data payload and transmits them in $r$ different slots of the frame. 
Different strategies for replica placement in the frame have been proposed in \cite{ValChiPao:22}.
The availability of a \ac{BS} with a massive number of antenna elements is a key feature to enable \ac{MPR} at the receiver. 
In addition to the massive number of \ac{BS} antennas, \ac{MPR} is enabled by the use of orthogonal preamble (or pilot) sequences.
In \ac{MMA} $K$ is typically much larger than the number of available pilots $N_\mathrm{P}$, so that pre-assignment of a specific orthogonal pilot to each user is not possible. 
As a strategy to cope with this issue, each active user picks one pilot randomly from the set of $N_\mathrm{P}$ available preambles, without any coordination with the other active users.
In this setting, if a user is the only one picking a particular pilot in a slot, decoding of the user message in that slot may succeed. 
Otherwise, under power control decoding fails and the \ac{SIC} procedure will be in charge of resolving the ``collision''.
The use of \ac{CSA}-based access and random pilot selection was proposed in \cite{Sor2018:coded}.

Regarding the channel model, we consider a block Rayleigh fading channel with \ac{AWGN}. 
The channel coherence time is assumed equal to the slot duration $T_\mathrm{s}$, which implies statistical independence of the channel coefficients of the same user across different slots. 
When coherence times are large, it is possible to subdivide slots into sub-slots as done recently in \cite{Liu22:unsourced} (where compressed sensing and \ac{SIC} across sub-slots are used), with the advantage that the user channel remains the same in all sub-slots. 
In this paper we consider relatively small coherence times and for this reason we stick with the framed and slotted structure.
We do not consider shadowing effects owing to the assumption of perfect power control.
Coherently with the above-mentioned access protocol and use of orthogonal pilots, each user active in a slot transmits a packet replica composed of one of the $N_\mathrm{P}$ orthogonal pilot sequences, of length  $N_\mathrm{P}$ symbols, concatenated with a data payload of length $N_\mathrm{D}$ symbols. 
Denoting the number of \ac{BS} antennas by $M$, the signal received in a slot may be expressed as $[\M{P}, \M{Y}] \in \mathbb{C}^{M \times (N_\mathrm{P}+N_\mathrm{D})}$ where
\begin{equation}\label{eq:P}
\begin{aligned}
    \M{P} &= \sum_{k \in \mathcal{A}} \V{h}_k \V{s}(k) + \M{Z}_p \\ 
    \M{Y} &= \sum_{k \in \mathcal{A}} \V{h}_k \V{x}(k) + \M{Z} .
\end{aligned}
\end{equation}
In \eqref{eq:P}, $\mathcal{A}$ is the set of users  transmitting a replica in the considered slot, while $\V{h}_k = (h_{k,1}, \dots, h_{k,M})^T \in \mathbb{C}^{M\times 1}$ is the vector of channel coefficients of the $k$-th user. 
The elements of $\V{h}_k$ are modeled as zero-mean, circularly symmetric, complex Gaussian \ac{i.i.d.} random variables, i.e., $h_{k,i} \sim \mathcal{CN}(0, \sigma_\mathrm{h}^2)$ for all $k \in \mathcal{A}$ and $i \in \{1,\dots,M\}$. 
Moreover, $\V{s}(k) \in \mathbb{C}^{1\times N_\mathrm{P}}$ and $\V{x}(k) \in \mathbb{C}^{1\times N_\mathrm{D}}$ are the orthogonal pilot sequence picked by user $k$ in the current slot and the user's payload, respectively, both with a unitary average energy per symbol.
Finally, $\M{Z}_p \in \mathbb{C}^{M\times N_\mathrm{P}}$ and $\M{Z} \in \mathbb{C}^{M\times N_\mathrm{D}}$ are matrices whose elements are Gaussian noise samples. 
The elements of both $\M{Z}_p$ and $\M{Z}$ are \ac{i.i.d.} random variables with distribution $\mathcal{CN}(0, \sigma_\mathrm{n}^2)$.
Due to power control, through the paper we adopt the normalization $\sigma_\mathrm{h}^2 = 1$ for all users' channel coefficients.

\subsection{Channel and Payload Estimation} \label{subsec:MRC}

As mentioned above, the \ac{BS} receives a signal in the form $[\M{P}, \M{Y}]$ in each slot of the frame. The processing can be split into two phases \cite{Sor2018:coded, ValChiPao:22}. In the first one, the \ac{BS} attempts channel estimation for all possible pilots by computing $\V{\phi}_j\in \mathbb{C}^{M\times 1}$, for all $j \in \{1,\dots,N_{\mathrm{P}}\}$, as
\begingroup
\allowdisplaybreaks
\begin{align}
\label{eq:PhiEstimate}
    \V{\phi}_j &= \frac{\M{P} \,\V{s}_j^{H}}{\| \V{s}_j \|^2} = \sum_{k \in \mathcal{A}^j} \V{h}_k + \V{z}_j
\end{align}
\endgroup
where $\mathcal{A}^j$ is the set of active devices employing pilot $j$ in the current slot, $\V{s}_j \in \mathbb{C}^{1\times N_\mathrm{P}}$ is the $j$-th pilot sequence, and $\V{z}_j \in \mathbb{C}^{M \times 1}$ is a noise vector with \ac{i.i.d.} $\mathcal{CN}(0, \sigma_\mathrm{n}^2 / N_\mathrm{P})$ entries.
Note that in absence of noise, when pilot $j$ is picked by a single user in the current slot, $\V{\phi}_j$ equals the vector of channel coefficients for that user.

In the second phase, the \ac{BS} computes the quantities $\V{f}_j \in \mathbb{C}^{1 \times N_\mathrm{D}}$ and $g_j \in \mathbb{R}$ as
\begingroup
\allowdisplaybreaks
\begin{align}
\label{eq:fComplete}
    \V{f}_j &= \V{\phi}_j^{H} \, \M{Y} \nonumber \\
    &= \sum_{k \in \mathcal{A}^j} \| \V{h}_k \|^2 \V{x}(k) + \sum_{k \in \mathcal{A}^j} \sum_{m \in \mathcal{A} \backslash \{k\}} \V{h}_k^H\,\V{h}_m \V{x}(m) + \V{\tilde{z}}_j
\end{align}
\endgroup
and
\begingroup
\allowdisplaybreaks
\begin{align}
\label{eq:gComplete}
    g_j &= \| \V{\phi}_j \|^2 \nonumber \\
    &= \sum_{k \in \mathcal{A}^j} \Bigg( \| \V{h}_k \|^2 + \sum_{m \in \mathcal{A}^j \backslash \{k\}} \V{h}_m^H\,\V{h}_k \Bigg) + \tilde{n}_j
\end{align}
\endgroup
where $\V{\tilde{z}}_j \in \mathbb{C}^{1 \times N_\mathrm{D}}$ and $\tilde{n}_j$ are noise terms. 
Then, the \ac{BS} attempts estimation of the payload using conventional \ac{MRC} as
\begin{align}
\label{eq:PayloadEst}
    \hat{\V{x}} = \frac{\V{f}_j}{g_j} = \frac{\V{\phi}_j^{H} \, \M{Y}}{\| \V{\phi}_j \|^2}\,.
\end{align}
In the case where a generic user $\ell$ is the only one transmitting with pilot $j$ in a given slot, hereafter referred to as singleton user ($\mathcal{A}^j = \{ \ell \}$), we have $\hat{\V{x}} \approx \V{x}_\ell$. 
Demapping and decoding operations are performed on $\hat{\V{x}}$ and, upon successful channel decoding, the packet symbols are stored in a buffer waiting for the \acl{SIC} phase. 
The aim of this latter iterative processing, that will be explained in detail in the next section, is to subtract the interference of a packet in a slot using the information retrieved in another slot from one of its replicas. 
In fact, whenever a packet is successfully decoded, the \ac{BS} acquires information about the positions of its replicas along with the employed preambles. 
This can be implemented in several ways, e.g., letting this information be a function of the information bits.
This information can be used to cancel interference from a slot and attempt the decoding procedure again.
Here, we separately computed $\V{f}_j$ and $g_j$ for reasons that will be clear in Section~\ref{subsec:NormSquared}.


\section{Analysis of \acl{SIC} Techniques} 
\label{sec:ImprovingSIC}

In this section we present our main contributions. 
We first review in detail a state-of-the-art \ac{SIC} technique for \ac{CSA} with massive \ac{MIMO} \cite{Sor2018:coded}, 
discussing some critical points. 
Then, we present a theoretical analysis of this technique to assess and investigate the role of interference. 
Motivated by the carried out analysis, we propose a \ac{SIC} algorithm to improve the overall \ac{CSA} performance.


\subsection{Channel Hardening-Based Interference Cancellation}\label{subsec:NormSquared}
Consider the low-complexity \ac{SIC} algorithm, here indicated as \ac{CHB}, proposed in \cite{Sor2018:coded} and also recently exploited in \cite{ValChiPao:22}. 
This algorithm relies heavily on channel hardening and favorable propagation effects, which hold when the number of \ac{BS} antennas, $M$, is large \cite{Bjo2017:MIMObook}. 
Accordingly, in a massive \ac{MIMO} setting, \eqref{eq:fComplete} and \eqref{eq:gComplete} can be approximated as
\begin{align}
\label{eq:fApprox}
    \V{f}_j &\approx \sum_{k \in \mathcal{A}^j} \| \V{h}_k \|^2 \V{x}(k) + \V{\tilde{z}} \\
\label{eq:gApprox}
    g_j &\approx \sum_{k \in \mathcal{A}^j} \| \V{h}_k \|^2 + \tilde{n}
\end{align}
respectively. 
In other words, the algorithm relies on assuming that the cross-terms in \eqref{eq:fComplete} and \eqref{eq:gComplete} (i.e., terms featuring a product $\V{h}_k^H\,\V{h}_m$ with $k \neq m$) can be neglected with respect to the main terms.
Assume that we initially compute $\V{f}_{j}$ and $g_j$, $j = 1, \dots, N_\mathrm{P}$, in all slots and that the payload of user $\ell$ is successfully decoded in a slot. Then, the above approximations lead naturally to the \ac{SIC} procedure where we update $\V{f}_j$ and $g_j$ as ${\V{f}_j} \leftarrow {\V{f}_j} - \| \V{h}_\ell \|^2\, \V{x}(\ell)$ and ${g_j} \leftarrow {g_j} - \| \V{h}_\ell \|^2$, respectively, in all slots where replicas of the $\ell$-th user's payload are present.
As such, this \ac{SIC} algorithm subtracts only the main interfering term from \eqref{eq:fComplete} and \eqref{eq:gComplete}.
The update requires knowledge of $\| \V{h}_\ell \|^2$ in the replica slots where, due to the block fading assumption, the channel coefficients are different. 
For this issue, in \cite{Sor2018:coded} the authors invoke temporal stability of $\| \V{h}_\ell \|^2$ through the whole frame. 
Here, we simply use the expectation $\E{\| \V{h}_\ell \|^2} = M$ to perform \ac{SIC} which is more accurate under block Rayleigh fading assumptions with $\sigma_\mathrm{h}^2 = 1$. 
Hence, the \ac{SIC} procedure can be described by the updates
\begin{align}\label{eq:CHB_SIS}
    {\V{f}_j} \leftarrow {\V{f}_j} - M\, \V{x}(\ell) \quad \text{and} \quad {g_j} \leftarrow {g_j} - M.
\end{align}

Before the next section, we want to foreshadow that the approximations \eqref{eq:fApprox} and \eqref{eq:gApprox} are not very accurate when the cardinality of $\mathcal{A}$ is large. 
In fact, since for $m \neq k$ we have
\begin{equation}
\begin{aligned}
    \E{\V{h}_k^H\,\V{h}_m} &= 0 \\ 
    \Var{\V{h}_k^H\,\V{h}_m} &= M
\end{aligned}
\end{equation}
the corresponding interfering terms in \eqref{eq:fComplete} and \eqref{eq:gComplete} may prevent from decoding a user packet even if it is the only one with a specific pilot. 
In the following we analyze this phenomenon by evaluating the probability that a user, being the only one with a specific pilot in a slot, is nevertheless not decoded.

\subsection{Theoretical Analysis of the Interference Effects}

\label{subsec:DecSISNS}
We use the terminology ``logical'' to refer to an idealized setting in which: (i) whenever a user is the only one using a pilot in a given slot it is successfully decoded with probability one;
(ii) channel estimation is perfect so that interference subtraction is ideal.
Hereafter we provide a theoretical analysis of the effects of interference by removing hypotheses (i) and (ii), to understand their impact in a realistic setting. 

Let us consider a situation where $|\mathcal{A}|$ users transmit simultaneously in a slot, $|\mathcal{A}^j|$ of them using pilot $j$.
Assume $|\mathcal{A}^j| - 1$ users from the set $\mathcal{A}^j$ have been successfully decoded in other slots of the frame. 
Then, in the current slot, we can apply \ac{CHB} interference subtraction which, as mentioned above, mitigate but does not eliminate completely the interference. At this point, there is only one undecoded user adopting the $j$-th pilot (singleton). 
To analyze the probability that this user is successfully decoded, we focus on the interfering and noisy terms in \eqref{eq:fComplete}. Then, from \eqref{eq:fComplete} we can write
\begingroup
\allowdisplaybreaks
\begin{align}
\label{eq:fTerms}
    \V{f}_j &= \sum_{k \in \mathcal{A}^j} \| \V{h}_k \|^2 \, \V{x}(k) + \V{I}_j
\end{align}
\endgroup
where
\begin{align}
\label{eq:I_j}
    \V{I}_j 
    &= \sum_{k \in \mathcal{A}^j} \sum_{m \in \mathcal{A} \backslash \{k\}} \V{h}_k^H\,\V{h}_m \,\V{x}(m) + \sum_{m \in \mathcal{A}} \V{z}_j^H \, \V{h}_m \,\V{x}(m) \nonumber \\ & 
    + \sum_{k \in \mathcal{A}^j} \V{h}_k^H \, \M{Z} + \sum_{m \in \mathcal{A}} \V{z}_j^H \, \M{Z}\,.
\end{align}
Let us define $\V{\xi}_1(k,m) = \V{h}_k^H\,\V{h}_m \,\V{x}(m)$. 
Since $\V{h}_k$ and $\V{h}_m$ are length-$M$ vectors whose entries are modeled as \ac{i.i.d.} $\mathcal{CN}(0, 1)$ random variables
and $\V{x}$ is a length-$N_\mathrm{D}$ payload vector with \ac{i.i.d.} entries, it follows that each entry $\xi_1(k, m)$ of $\V{\xi}_1(k,m)$, $k \neq m$, fulfills
\begin{equation}
\label{eq:Xi1MeanVar}
\begin{aligned}
    \E{\xi_1(k,m)} &= 0 \\ 
    \Var{\xi_1(k,m)} &= M \,.
\end{aligned}
\end{equation}
The second group of terms in \eqref{eq:I_j} can be represented by $\V{\xi}_2(m) = \V{z}_j^H \, \V{h}_m \,\V{x}(m)$ where  $\V{z}_j$ is a noise vector with \ac{i.i.d.} $\mathcal{CN}(0, \sigma_\mathrm{n}^2 / N_\mathrm{P})$ entries. 
Therefore each entry $\xi_2(m)$ of $\V{\xi}_2(m)$ fulfills
\begin{equation}
\label{eq:Xi2MeanVar}
\begin{aligned}
    \E{\xi_2(m)} &= 0 \\ 
    \Var{\xi_2(m)} &= \frac{M}{N_\mathrm{P}}\,\sigma_\mathrm{n}^2 \,.
\end{aligned}
\end{equation}
Similarly, the third group of terms in \eqref{eq:I_j} can be represented by $\V{\xi}_3(k) = \V{h}_k^H \, \M{Z}$ where  
$\M{Z}$ is a matrix whose elements are \ac{i.i.d.} $\mathcal{CN}(0, \sigma_\mathrm{n}^2)$. 
Then, each entry $\xi_3(k)$ of $\V{\xi}_3(k)$ fulfills
\begin{equation}
\begin{aligned}
\label{eq:Xi3MeanVar}
    \E{\xi_3(k)} &= 0 \\ 
    \Var{\xi_3(k)} &= M\,\sigma_\mathrm{n}^2 \,.
\end{aligned}
\end{equation}
Finally the last term $\V{\xi}_4 =\V{z}_j^H \, \M{Z}$ has entries characterized by
\begin{equation}
\label{eq:Xi4MeanVar}
\begin{aligned}
    \E{\xi_4} &= 0\\ 
    \Var{\xi_4} &= \frac{M}{N_\mathrm{P}}\,\sigma_\mathrm{n}^4 \,.
\end{aligned}
\end{equation}

We can now make the approximation which considers entry independence between $\V{\xi}_1(k,m)$, $\V{\xi}_2(m)$,  $\V{\xi}_3(k)$, and $\V{\xi}_4$. 
Under this approximation it follows that the entries ${I}_j$ of $\V{I}_j$ have
\begin{equation}
\label{eq:I_j_MeanVar}
\begin{aligned}
    \E{{I}_j} &= 0 \\ 
    \Var{{I}_j} &= M\left( |\mathcal{A}^j| \left(|\mathcal{A}| - 1 + \sigma_\mathrm{n}^2\right) + \frac{\sigma_\mathrm{n}^2}{N_\mathrm{P}} \left(|\mathcal{A}| + \sigma_\mathrm{n}^2\right)\right)\,.
\end{aligned}
\end{equation}
As mentioned above, let $|\mathcal{A}^j| - 1$ users employing pilot $j$ be decoded in other slots. 
Performing \ac{CHB} \ac{SIC} \eqref{eq:CHB_SIS}, new residual interfering terms arise. 
Equation \eqref{eq:fTerms} can be rewritten as
\begin{align}
\label{eq:fTerms2}
    \V{f}_j &= \| \V{h}_\ell \|^2\, \V{x}(\ell) + \sum_{k \in \mathcal{A}^j \backslash \{ \ell\}} \left(\| \V{h}_k \|^2 - M\right)\, \V{x}(k) + \V{I}_j \nonumber \\
    &= \| \V{h}_\ell \|^2\, \V{x}(\ell) + \tilde{\V{I}}_j
\end{align}
where the subscript $\ell$ denotes the only remaining user employing pilot $j$ in the slot.
Since $\E{\| \V{h}_k \|^2} = M$ and $\Var{\| \V{h}_k \|^2} = M$, we can incorporate these terms in our approximation, leading to
\begin{equation}
\begin{aligned}
\label{eq:Itilde_j_MeanVar}
    \E{\tilde{I}_j} &= 0 \\ 
    \Var{\tilde{I}_j} &= M\left( |\mathcal{A}^j| \left(|\mathcal{A}| + \sigma_\mathrm{n}^2\right) - 1 + \frac{\sigma_\mathrm{n}^2}{N_\mathrm{P}} \left(|\mathcal{A}| + \sigma_\mathrm{n}^2\right)\right)\,.
\end{aligned}
\end{equation}
Due to summation of a large amount of terms, we can approximate $\tilde{I}_j$ as a circularly symmetric complex Gaussian distribution with the mean and variance given in \eqref{eq:Itilde_j_MeanVar}.
Then, dividing by $M$ we can estimate the payload of user $\ell$ as
\begin{align}
\label{eq:EqChModel}
    \hat{\V{x}}(\ell) = \frac{\|\V{h}_\ell\|^2}{M} \, \V{x}(\ell) + \frac{\tilde{\V{I}}_j}{M} \,.
\end{align}

For a realistic analysis we also consider modulation and channel coding. 
Employing an $\mathsf{M}$-\ac{QAM} modulation and hard-decision decoding, the symbol error probability for given $w = \frac{2}{\sigma^2_h} \, \| \V{h}_\ell \|^2$ can be written as \cite{Con2005:MQAM}
\begin{align}\label{eq:peGivenHMQAM}
    P_ {\mathrm{e}|w} 
    &=  A_\mathsf{M}\,\text{erfc}\left(\sqrt{\frac{C_\mathsf{M} \, w^2}{\Var{\tilde{I}_j} }}\right) - \frac{A_\mathsf{M}^2}{4}\, \text{erfc}^2\left(\sqrt{\frac{C_\mathsf{M}\,w^2}{\Var{\tilde{I}_j}}}\right)
\end{align}
where $A_\mathsf{M} = 2-2/\sqrt{\mathsf{M}}$ and $C_\mathsf{M} = 3/ (8\mathsf{M}-8)$.
Finally, we assume an error correcting code with bounded-distance hard-decision decoding, able to correct up to $t$ errors, and constellation Gray mapping.
We can express the probability that decoding of a user packet is unsuccessful given $w$ as
\begin{align}
\label{eq:PfailGivenw}
    P_{\mathrm{fail}|w} \approx 1 - \sum_{d = 0}^{t} \binom{N_\mathrm{D}}{d}\, P_{\mathrm{e}|w}^d \left( 1 - P_{\mathrm{e}|w} \right)^{N_\mathrm{D}-d}
\end{align}
where $N_D$ is the number of payload symbols.
Equality in \eqref{eq:PfailGivenw} would hold if, whenever a symbol is failed, only one of its bits was received in error. 
In general this is not true, but exploiting Gray mapping this is a well-fitting approximation.

In conclusion, under \ac{CHB} \ac{SIC}, the probability that decoding of a user packet is unsuccessful in a slot where its $|\mathcal{A}^j| - 1$ pilot-interferers have been subtracted and a total of $|\mathcal{A}|$ users were initially allocated in the slot can be expressed as
\begin{align}
\label{eq:Pfail}
    P_\mathrm{fail} = \int_{0}^{\infty} P_{\mathrm{fail}|w} \, \frac{1}{2^M\, \Gamma(M)} \, w^{M-1}\, e^{-w/2}\,dw
\end{align}
where $\Gamma(\cdot)$ is the gamma function.
This follows from $w$ being chi-squared distributed with $2M$ degrees of freedom ($\sigma^2_h = 1$). 
The expression assumes $\mathsf{M}$-\ac{QAM} constellation with Gray mapping and hard-decision decoding.
The expression of $P_{\mathrm{fail}|w}$ in \eqref{eq:Pfail} is given by \eqref{eq:PfailGivenw}, where $P_{\mathrm{e}|w}$ is provided in \eqref{eq:peGivenHMQAM} with the approximation \eqref{eq:Itilde_j_MeanVar}.
We observe that, to increase the resilience of singleton users to interference in terms of packet error probability, we can increase either the number of \ac{BS} antennas $M$ or the code error correction capability $t$ for fixed $N_\mathrm{D}$ (which however decreases the error correcting code rate). In Appendix~\ref{app:GeneralizationModCod}, we show how to extend this analysis to general modulation and coding schemes.%

\begin{remark}
A simple approximation can be made, observing that for large $M$ the \ac{PDF} narrowed around the mean value and therefore $\E{f(w)} \simeq f(\E{w})$. This was made in \cite{Valentini2022:Impact} neglecting the noise (i.e., $\sigma_\mathrm{n}^2 = 0$). 
\end{remark}

\begin{figure}[t]
    \centering
    \includegraphics[width=0.99\columnwidth]{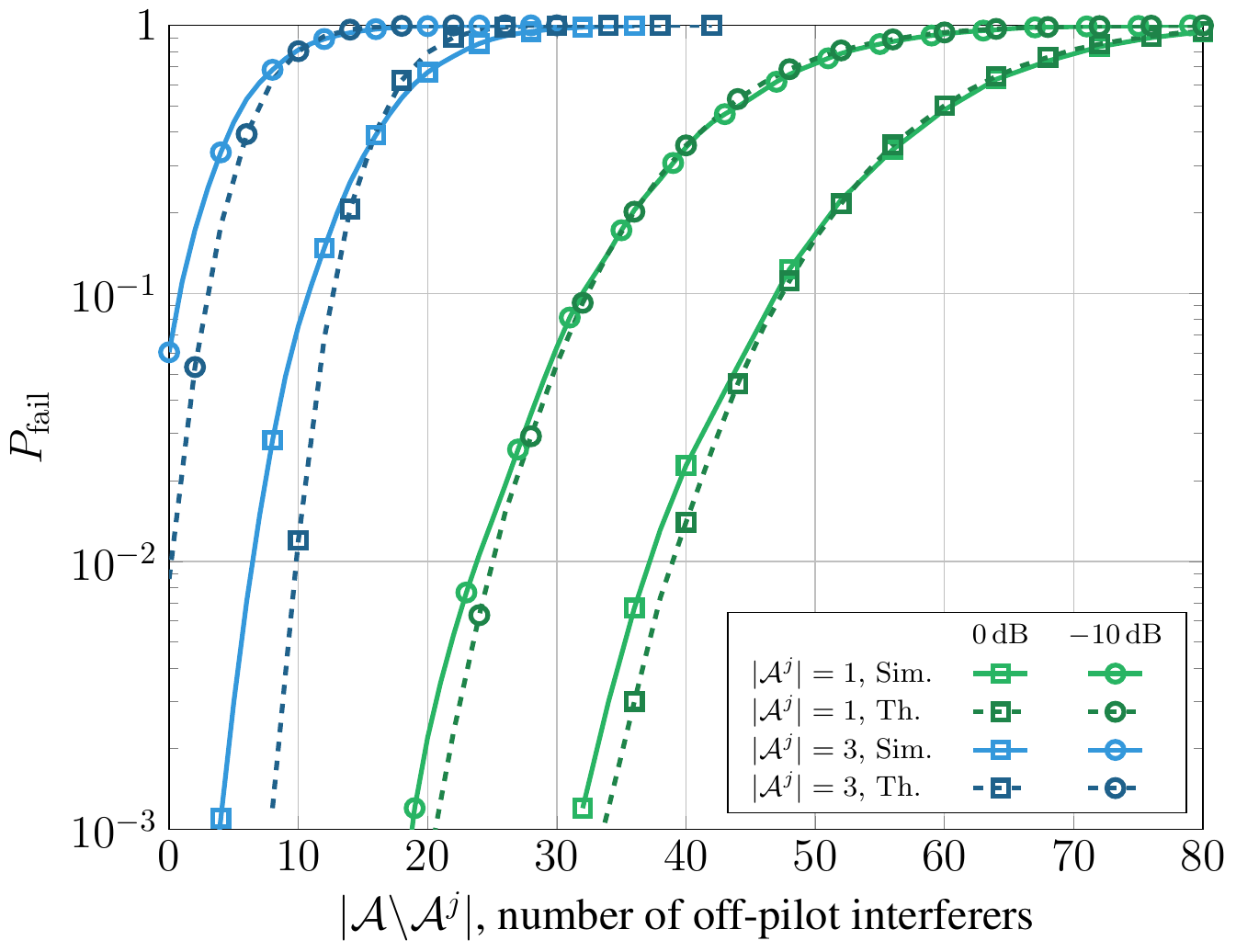}
    \caption{Probability of decoding failure of a singleton user after $|\mathcal{A}^j| - 1$ \ac{CHB} interference subtraction operations. Comparison between the analytical approximation \eqref{eq:Pfail} and the simulation for $N_\mathrm{D} = 256$, $t = 10$, $M = 256$, \ac{QPSK} constellation, and $\sigma_\mathrm{n}^2 \in \{ 1, 10\}$.}
    \label{fig:InterfNS_Noise}
\end{figure}

\begin{example}
We report in Fig.~\ref{fig:InterfNS_Noise} the analytical approximations derived in \eqref{eq:Pfail} in comparison with Monte Carlo simulations for $N_\mathrm{D} = 256$, $t = 10$, $M = 256$, \ac{QPSK} constellation, and two noise levels $\sigma_\mathrm{n}^2 \in \{ 1, 10\}$.
Despite approximations, the analytical results provide a good estimate of the simulated curves also in the presence of noise. 
In particular, when $|\mathcal{A}^j| = 1$, no interference subtractions are performed and the user experiences the most favorable interference conditions. 
The $|\mathcal{A}^j| = 1$ curve in Fig.~\ref{fig:InterfNS_Noise} reveals the actual performance of \ac{MRC} payload estimation in \eqref{eq:PayloadEst} when interferers, using different orthogonal preambles, are captured in the model.
Indeed, this is a major non-ideality, degrading the general performance of \ac{MAC} protocols when a realistic channel model is considered.
On the other hand, when $|\mathcal{A}^j| > 1$, the estimation deteriorates even more, revealing the non-ideality of the \ac{SIC} procedure. 
Moreover, we point out that, whenever a device using pilot $j$ in the current slot is successfully decoded and \ac{CHB} is performed, the interference on pilots different from $j$ is not mitigated.
This is the critical point of this \ac{SIC} procedure and in Section~\ref{subsec:PayloadAided} we will propose a technique able to overcome this problem.
\end{example}

\begin{remark}
The analysis conducted in this section, not only provides system design guidelines, but can be relevant to jointly optimize \ac{PHY} and \ac{MAC} layer. 
For example, in \cite{Valentini2022:Joint} an optimization is proposed based on density evolution recursion under realistic channel and \ac{PHY} layer processing relying on \eqref{eq:Pfail}.
\end{remark}

\subsection{Payload Aided Subtractions}\label{subsec:PayloadAided}
Motivated by the analysis carried out in the previous subsection, we aim at changing the \ac{SIC} algorithm to make it more effective and improve the overall performance.
In repetition-based \ac{CSA}, users send multiple copies of the same payload over the frame.
Hereafter, we refer to the slots in which a packet is successfully decoded as ``generator'' slots. 

Assume one of the replicas sent by a user, say user $\ell$, is successfully decoded in a slot, in correspondence of some pilot $\V{s}_j$. 
The \ac{BS} available information consists of the user's payload $\V{x}(\ell)$, which is common to all replicas, the indexes of the slots where the other replicas have been transmitted, the indexes of the pilots used in each such replica, and the estimate $\V{\phi}_j$ of the channel coefficients in the generator slot computed as per \eqref{eq:PhiEstimate}.
The interference subtraction operation in the generator slot is performed as
\begin{equation}
\begin{aligned}\label{eq:PYmod}
\M{P}^{(i+1)} &= \M{P}^{(i)} - \V{\phi}_j \V{s}_j \\
\M{Y}^{(i+1)} &= \M{Y}^{(i)} - \V{\phi}_j \V{x}(\ell)
\end{aligned}
\end{equation}
where we let $\M{P}^{(0)} = \M{P}$ and $\M{Y}^{(0)} = \M{Y}$. 
As from \eqref{eq:PYmod}, in the generator slot we do not recompute the channel estimate since the estimation provided by $\V{\phi}_j$ is impaired only by noise.
Regarding the replica slots, we exploit knowledge of the payload (that is the same in all replicas) to estimate the channel coefficients as
\begingroup
\allowdisplaybreaks
\begin{align}
\label{eq:hTilde}
    \hat{\V{h}}^{(i)}_\ell &= \frac{\M{Y}^{(i)} \,\V{x}(\ell)^{H}}{\| \V{x}(\ell) \|^2} = \V{h}_\ell + \tilde{\V{h}}_\ell\,.
\end{align}
\endgroup
Then, using the ``payload-based'' channel estimate, in the replica slots we can perform subtraction of interference, similar to \eqref{eq:PYmod}, as
\begin{equation}
\begin{aligned}\label{eq:PYmodTilde}
    \M{P}^{(i+1)} &= \M{P}^{(i)} - \hat{\V{h}}^{(i)}_\ell \V{s}(\ell) \\
    \M{Y}^{(i+1)} &= \M{Y}^{(i)} - \hat{\V{h}}^{(i)}_\ell \V{x}(\ell)\,.
\end{aligned}
\end{equation}
In this \ac{SIC} algorithm, hereafter referred to as \ac{PAB}, each time an update of the matrices $\M{P}$ and $\M{Y}$ has been carried out we re-compute \eqref{eq:PhiEstimate} and \eqref{eq:PayloadEst} for each pilot in the current slot, to check if any other user can be successfully decoded after interference subtraction. 
We point out that exploiting the preamble (instead of the payload) to perform channel estimation in slots where we wish to subtract interference may heavily deteriorate the estimation quality due to preamble collisions. 
For the sake of clarity, we report in Algorithm~\ref{alg:AlgoSummary} the base station \ac{SIC} processing.

\begin{algorithm}[t]
\SetNlSty{textbf}{}{:}
\While{``decoded user buffer is not empty''}{
	pop user $\ell$ information from the buffer\;
	\For{``all replicas of user $\ell$''}{
        res = slot-pilot pair of the current replica of $\ell$\;
  		\uIf{SIC == CHB}{
   			Use (8) and re-attempt decoding in res\;
        }\Else(\tcp*[h]{SIC == PAB}){
            \uIf{``res is the resource where the user was found''}{
                Use (23) in the slot of res\;
            }\Else{
                Re-estimate channel with (24) and use (25)\; 
            }
            Re-attempt decoding for all pilots in the slot of res\;
        } 
        Update the buffer with new found users if necessary\;
	}
}
\caption{Base station SIC processing summary.}
\label{alg:AlgoSummary}	
\end{algorithm}

\begin{remark}
In the particular case in which we perform 
the first subtraction operation in a slot using \eqref{eq:PYmodTilde}, we have
\begingroup
\allowdisplaybreaks
\begin{align}
\label{eq:hTildeExample}
    \hat{\V{h}}^{(0)}_\ell &= \V{h}_\ell + \sum_{k \in \mathcal{A} \backslash \{\ell\}} \V{h}_k \frac{\V{x}(k) \,\V{x}(\ell)^{H}}{\| \V{x}(\ell) \|^2} + \V{z}_\mathrm{h}
\end{align}
\endgroup
where $\V{z}_\mathrm{h}$ is the residual noise term. 
In this specific case, we can derive the statistical properties of the estimation error $\tilde{\V{h}}_\ell$, given that the payload symbols are independent among users, as
\begin{equation}
\begin{aligned}
    \E{\tilde{{h}}_{\ell,n}} &= 0 \\  \Var{\tilde{{h}}_{\ell,n}} &= \frac{|\mathcal{A}| - 1 + \sigma_n^2}{N_\mathrm{D}}
    \label{eq:VarHTilde}
\end{aligned}
\end{equation}
where $n = 1, \dots, M$.
We observe that, as expected, the accuracy of the channel coefficients estimate improves as the number of payload symbols increases. 
On the other hand, the channel estimate deteriorates as the number of users transmitting in the slot increases. 
Among all possible $\tilde{\V{h}}_\ell$ obtained running the \ac{SIC} algorithm, this represents the worst case 
in terms of estimation accuracy.
\end{remark}

\begin{figure}[t]
    \centering
    \includegraphics[width=0.99\columnwidth]{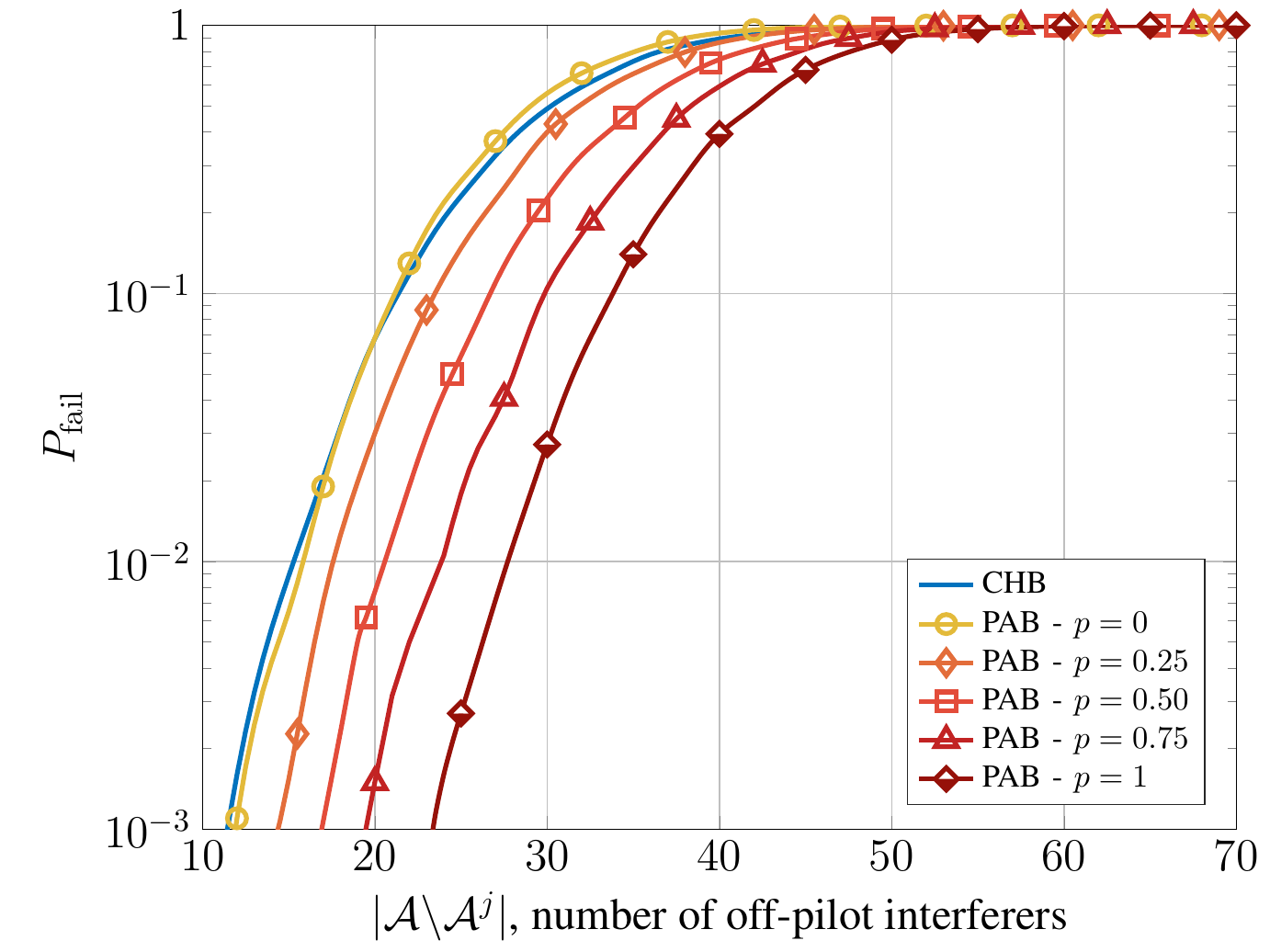}
    \caption{Probability of decoding failure of a singleton user after one interference subtraction operation ($|\mathcal{A}^j| = 2$). Comparison between \ac{CHB} and PAB for $N_\mathrm{D} = 256$, $t = 10$, $M = 256$, and $\sigma_\mathrm{n}^2 = 0$.}
    \label{fig:InterfhEst}
\end{figure}

In general, at step $i = n_\mathrm{up} + n_\mathrm{pa}$ of the \ac{PAB} \ac{SIC} algorithm, $n_\mathrm{up}$ subtractions using \eqref{eq:PYmod} have been performed (due to uncollided pilots), and $n_\mathrm{pa}$ ones based on payload-aided channel coefficients estimation as from \eqref{eq:PYmodTilde}.
In this regard, Fig.~\ref{fig:InterfhEst} illustrates the results of a variation of the experiment described in Section~\ref{subsec:DecSISNS} for the two \ac{SIC} techniques, \ac{CHB} and \ac{PAB}, and for $|\mathcal{A}^j| = 2$. 
More specifically, we assume that a fraction $0 \le p \le 1$ of users in the set $\mathcal{A} \backslash \mathcal{A}^j$ have been successfully decoded and subtracted. For them, we consider the worst case scenario ($n_\mathrm{up} = 0$) where the \ac{SIC} is performed using \eqref{eq:PYmodTilde}.
As expected, the \ac{PAB} performance improves as $p$ increases. 
On the other hand, \ac{CHB} is not influenced by $p$. Then, averaging on $p$, \ac{PAB} outperforms \ac{CHB} also in the worst case scenario.
Moreover, as in a real scenario we have $n_\mathrm{up} > 0$, the \ac{PAB} technique is expected to outperform the \ac{CHB} one to a larger extent; this is confirmed by the numerical results presented in Section~\ref{sec:NumericalResults}.

\begin{remark}
A variation of the \ac{PAB} technique can be applied also in case of protocols featuring payload segmentation and packet-level coding applied to the segments \cite{paolini2015:csa}.
When a sufficient number of segments have been decoded in their slots, packet-level decoding allows reconstructing the missing segments.
To subtract the interference generated by the missing segments in their slots, ``segment-based'' channel estimation can be performed, similar to payload-based one when payloads are replicated.
\end{remark}

\subsection{Scheduling of Interference Subtraction Operations}

In this section we propose a processing technique that is able to enhance the overall system performance and that can be applied in different \ac{MAC} and \ac{PHY} layer configurations. 
The approach consists of introducing a priority scheduler for interference subtraction operations based on the accuracy of the corresponding channel estimation.

Let us initially focus our attention on \ac{PAB} schemes. 
Recalling \eqref{eq:PhiEstimate}, we see that pilot-based channel estimation is impaired by noise only in case of a singleton user (namely, when $|\mathcal{A}^j| = 1$ for some $j$). 
On the other hand, the samples corresponding to replicas of a successfully decoded packet are subtracted from the received matrices $\M{P}$ and $\M{Y}$ using the payload-based estimation of the channel coefficient according to \eqref{eq:PYmodTilde}. 
Since the payloads are not orthogonal with each other, payload-based estimation is impaired by both noise and interference. 
We can therefore categorize interference subtraction operations based on the accuracy of the channel estimation on which they rely and schedule ``high quality'' subtractions first.
Since in each \ac{SIC} iteration channel estimations are performed on the current $\M{P}$ and $\M{Y}$ matrices, as per \eqref{eq:PhiEstimate} and \eqref{eq:hTilde}, it is expected that giving priority to those subtractions that deteriorate these matrices less (in terms of interference residue after the subtraction is performed) helps to increase the number of successful channel decoding operations, triggering further \ac{SIC} iterations and avoiding a premature stop of the \ac{SIC} process. 

The proposed scheduling of interference subtraction operations, hereafter referred to as ``instantaneous cancellation'', works as follows.
Consider phase 1 of the \ac{BS} processing.
After reception of each block of symbols corresponding to a slot, the \ac{BS} attempts packet decoding for each pilot by computing and processing $\hat{\V{x}}$ in \eqref{eq:PayloadEst}.
In the baseline scheduling described in literature (e.g., in \cite{Sor2018:coded}), all packets successfully decoded in the slot are buffered, awaiting for \ac{SIC} phase. 
Then, after all slots have been processed and the \ac{SIC} phase starts, the subtraction operations are scheduled following the order in which the decoded packets are extracted from the buffer (e.g., according to a first-in-first-out or last-in-first-out policy): for each extracted packet, the samples of all its replicas are subtracted in parallel from the corresponding slots.
Instead of this, we propose to perform subtraction operations of singleton users, i.e., high-quality subtraction operations relying on pilot-based channel estimates, immediately after a packet has been decoded in slot and to immediately reprocess the other pilots in the same slot, iterating the procedure and moving to the next slot only when no new packets can be successfully decoded.
The successfully decoded packets are still buffered awaiting for the \ac{SIC} phase, but the samples of these packets are ``instantaneously'' subtracted from the slots where they have been decoded.

\begin{figure}[t]
    \centering
    \includegraphics[width=0.9\columnwidth]{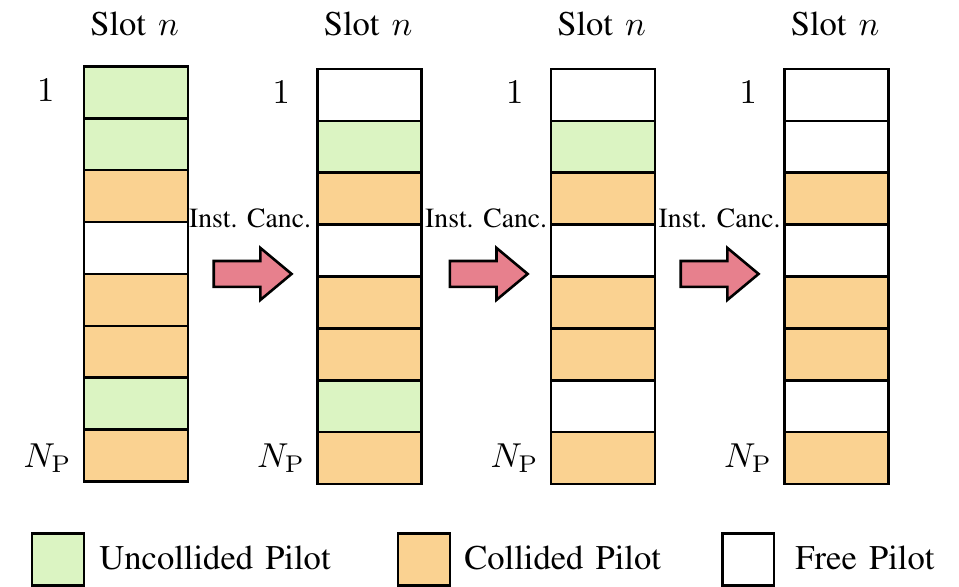}
    \caption{Pictorial representation of the instantaneous cancellation technique. In the example have been used $N_\mathrm{P} = 8$ orthogonal pilots per slot. In green are represented pilots chosen by one user (singleton), in orange the pilots used by two or more users, and in white the unused pilots.}
    \label{fig:ExampleInstaCanc}
\end{figure}

This provides a second benefit which is exemplified in Fig.~\ref{fig:ExampleInstaCanc}. 
In this example, the total number of pilots is $N_\mathrm{P} = 8$ and we are processing the generic slot $n$. 
There are three singleton users in pilot $p \in \{ 1, 2, 7\}$, pilot $4$ is unused, while the other pilots have been chosen by more than one user. 
Let pilots be considered in order from $1$ to $N_{\mathrm{P}}$, and assume the decoder successfully decodes a packet in correspondence of pilot $1$.
It performs instantaneous cancellation and re-attempts decoding from pilot $1$\footnote{Considering that only a singleton user can be successfully decoded, the procedure can be optimized avoiding to search for packets in pilots where a user has already been found.}.
Next, assume that when the receiver attempts decoding in pilot $2$, a decoding failure occurs. 
Such an event is consistent with the curves in Fig.~\ref{fig:InterfNS_Noise}, for $|\mathcal{A}^j| = 1$, which illustrate that even a singleton user may not been correctly decoded due to interference and noise. 
Then, let the receiver successfully decode a packet using pilot~$7$ (decoding failures necessarily occur in pilots from $3$ to $6$) and immediately subtract the corresponding samples from the slot: 
Since the receiver restarts again from pilot~$1$ and now in the slot there is less interference compared to the previous decoding step, it is possible that the packet using pilot $2$ is now decoded. 
Deferring all subtractions to the \ac{SIC} phase and performing in parallel all subtractions associated with the same packet, the user in pilot $2$ could not be found; even if the user was found in another slot, the subtraction in the slot $n$ would be impaired by both noise and interference, deteriorating the overall performance.

This algorithm synergizes effectively also with \ac{MAC} protocols that foresee a feedback channel used by the \ac{BS} to broadcast \ac{ACK} messages, e.g., at the end of each slot. 
This is because, when the scheduling algorithm is applied, a larger number of \ac{ACK} messages are more likely to be triggered. 
In general, the instantaneous cancellation technique can be seen as a pre-\ac{SIC} processing that is performed slot by slot and, as such, can be employed by both \ac{CHB} and \ac{PAB} processing schemes. 
In Section~\ref{sec:NumericalResults} we will show the effectiveness of this technique for different choices of the \ac{MAC} access protocol and \ac{PHY} layer processing.

\subsection{Complexity Analysis}

In this subsection, we discuss the \ac{BS} processing complexity. 
Firstly, we carry out a worst case complexity analysis, assuming that no particular strategy aimed at reducing the cost of processing is applied. 
Possible optimization techniques to lower complexity are pointed out at the end of the subsection.
As from Section~\ref{subsec:MRC}, we can split the \ac{BS} processing into two phases, the initialization one (i.e., slot-by-slot processing) and the \ac{SIC} one. 
Hereafter, we denote by $C_\mathrm{INIT}$ the cost of initializing one slot and $C_\mathrm{SIC}$ the cost of subtracting the inference of one user, such that the total cost is given by $C_\mathrm{TOT} = N_\mathrm{s}\,C_\mathrm{INIT} + K_\mathrm{a}\,C_\mathrm{SIC}$ (assuming all $K_{\mathrm{a}}$ users active in the frame are processed, otherwise the expression is an upper bound).

The typical situation is the one where the cost of channel decoding, here referred to as $C_\mathrm{DEC}$, dominates all the other costs involved in \eqref{eq:PhiEstimate}, \eqref{eq:PayloadEst}, \eqref{eq:CHB_SIS}, \eqref{eq:PYmod}, \eqref{eq:hTilde}, and \eqref{eq:PYmodTilde}, including matrix multiplications, matrix subtractions, and scalar divisions.
This is true not only for the here considered algebraic linear block codes with bounded-distance hard-decision decoding, but also for \ac{LDPC} codes under belief-propagation decoding or polar codes under successive cancellation list decoding.
Then, since during the initialization phase we attempt decoding $N_\mathrm{P} \beta$ times in each slot, where $\beta = 1$ when instantaneous cancellation is not applied and $\beta = N_\mathrm{P}$ (in the worst case) otherwise, we have $C_\mathrm{INIT} \approx N_\mathrm{P} \, \beta \, C_\mathrm{DEC}$.
Similarly, during the \ac{SIC} phase we perform decoding $\gamma \, r \alpha$ times per each interfering user, where $\gamma = 1$ when instantaneous cancellation is not applied and $\gamma \le 1$ otherwise, and where $\alpha = 1$ for \ac{CHB} and $\alpha = N_\mathrm{P}$ (in the worst case) for \ac{PAB}.
Here, $\alpha$ represents the average number of decoding re-attempts per slot, while $\gamma \, r$ may be regarded as the ``effective'' number of replicas to be subtracted per user in the \ac{SIC} phase.
This leads us to $C_\mathrm{SIC} \approx \gamma \, r  \, \alpha \, C_\mathrm{DEC}$.
We conclude that the total cost may be expressed as $C_\mathrm{TOT} \approx \left( N_\mathrm{s} N_\mathrm{P} \, \beta \, + K_\mathrm{a} \, \gamma \, r  \, \alpha \right)C_\mathrm{DEC}$. 
Comparing the total cost of the low-complexity \ac{CHB} scheme with that of \ac{PAB} with instantaneous cancellation (highest complexity) we see that, in this worst case analysis, the increase in complexity is linear by a factor of approximately $N_\mathrm{P}$.

Let us finally discuss how, in practice, the cost of \ac{PAB} with instantaneous cancellation can be significantly reduced.
Activity detection techniques (e.g., a simple energy detector) are effective in decreasing the value of $\alpha$ as they allow avoiding to attempt decoding on empty or too crowded slot-pilot pairs. 
The value of $\beta$ can be lowered in the same way by avoiding useless decoding attempts in initialization phase.
Also very simple (and easy to implement) tricks are effective to substantially reduce complexity. 
For example, simply avoiding to reprocess pilots where a user has already been found during instantaneous cancellation, it is possible to drop the value of $\beta$ from $N_\mathrm{P}$ to $(N_\mathrm{P}+1)/2$.
In general, by the means of such optimizations it is possible to reduce $\alpha$ and $\beta$ to values much smaller than $N_{\mathrm{P}}$ (usually between $1$ and $5$, depending on the traffic, with $N_{\mathrm{P}}=64$).
Lastly, it is very important to point out that the actual processing time is very dependent of the architecture: for example, the operations increasing complexity (i.e., decoding attempts) well-fit parallel computational architectures since are independent of each other.

\medskip


\subsection{Collision Channel Benchmarks}

In this section we introduce some performance benchmarks that will be used in Section~\ref{sec:NumericalResults}. These benchmarks are based on a collision channel model over ``resources'' (slot-pilot pairs), on a collision channel model without \ac{SIC}, and on a more realistic setting we name \ac{PRCE}, respectively. 
In addition, we provide the analytical expression for the no-\ac{SIC} performance, and show that the \ac{PRCE} benchmark is approachable under specific conditions.

The system performance assuming a collision channel over resources provides an upper bound on the number of simultaneously active users at a target reliability.
In this idealized setting: (i) a packet arriving alone in a slot-pilot pair is successfully decoded with probability one (meaning perfect channel estimation and very high signal-to-noise ratio);
(ii) interference cancellation in the generator slot and across slots is perfect (meaning perfect channel estimation for the replicas); (iii) no decoding is possible of multiple packets arriving in the same resource (typical in presence of power control).
This assumption can be seen as an extension of the classical collision channel over slots. 
When evaluating numerical results we refer to this benchmark as ``logical performance with \ac{SIC}''.

\begin{figure}[t]
    \centering
    \includegraphics[width=0.9\columnwidth]{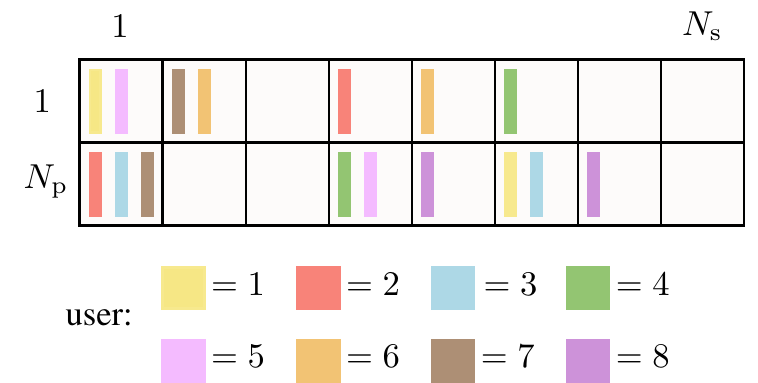}
    \caption{An example of user replicas allocation in a frame with $N_\mathrm{s} = 8$ slots and $N_\mathrm{P} = 2$ orthogonal pilots. There are $K_\mathrm{a} = 8$ active users, each of them transmitting $r = 2$ packets in the frame.}
    \label{fig:ExampleBenchmark}
\end{figure}

\begin{example} \label{ex:CollCh}
In Fig.~\ref{fig:ExampleBenchmark} we provide an example assuming allocation of the users' replicas  in a frame with $N_\mathrm{s} = 8$ slots and $N_\mathrm{P} = 2$ orthogonal pilots. 
There are $K_\mathrm{a} = 8$ active users, each of them transmitting $r = 2$ packets. 
We use the notation $(s, p)$ to indicate the resource corresponding to slot $s$ and pilot $p$.
Considering collision channel over resources, the messages of users $2$, $6$, $8$, and $4$ are successfully decoded in resources $(4, 1)$, $(5, 1)$, $(5, 2)$, and $(6, 1)$ respectively. 
Note that the message of user $8$ is decoded also in resource $(7, 2)$. 
Then, \ac{SIC} is performed for all decoded users, leaving user $7$ and $5$ in $(2, 1)$ and $(4, 2)$ uncollided. 
Iterating this procedure until no more packets are found, it is easy to verify that all users are retrieved in the order $2, 6, 8, 4, 7, 5, 1, 3$.
\end{example}

As a ``worst case'' benchmark, we consider also the situation where collision channel over resources model is adopted, but no \ac{SIC} procedure is run at the receiver. 
When replicas are randomly placed in the frame, the performance curve in terms of packet loss probability, given that there are $K_\mathrm{a}$ simultaneously active users, can be analytically derived (see Appendix~\ref{sec:WithoutSIC}) as
\begin{align}
	P_\mathrm{L, \,noSIC} &= \left(1 - \left(1 - \frac{r}{N_\mathrm{s}N_\mathrm{P}}\right)^{K_\mathrm{a}-1}\right)^r
\end{align}
for $N_\mathrm{s}$ slots per frame, $N_\mathrm{P}$ orthogonal pilots, and $r$ replicas per user. 
This analysis allows assessing the improvement on the massive access schemes attributable to the \ac{SIC} processing. 
When evaluating numerical results, we refer to this benchmark as ``logical performance without \ac{SIC}''.

\begin{example}
As reported in Example~\ref{ex:CollCh} initialization phase, only user $2$, $6$, $8$, and $4$ are retrieved referring to Fig.~\ref{fig:ExampleBenchmark}. Then, since no \ac{SIC} algorithm is considered, all the other user messages are lost.
\end{example}

As a third benchmark, we consider a more realistic setting (compared to collision channel assumptions) in which payload estimation is performed as in \eqref{eq:PayloadEst}; upon successful message decoding in a slot, \ac{PAB} processing is applied under the assumption that the subtractions are perfect (ideal \ac{SIC}). 
In this setting, referred to as as \ac{PRCE}, the performance is therefore limited by payload estimation \eqref{eq:PayloadEst} only. 
This establishes a second upper bound on the number of simultaneously active users; this upper bound is generally tighter than the logical performance with \ac{SIC} one.

\begin{example}
With reference again to Fig.~\ref{fig:ExampleBenchmark}, some of the replicas from users $2$, $6$, $8$, and $4$ are singleton ones in the corresponding resources. 
Under a collision channel over resources model, these replicas would be decoded with probability one.
However, since in the \ac{PRCE} setting payload estimation is realistic and might fail (as it was revealed in the analysis yielding Fig.~\ref{fig:InterfNS_Noise}), the process \ac{SIC} may stop prematurely. 
\end{example}

\begin{remark}
The \ac{PRCE} performance can be approached, under real channel estimation conditions, when the coherence time of the channel is larger than $r$ times the slot duration and we adopt the access protocol proposed in~\cite{ValChiPao:22} and called intra-frame \ac{SC}. 
This access strategy consists of letting each active device transmit its replicas in nearby slots, again with a random pilot selection for each replica.
In such a setting, in the high \ac{SNR} regime, the channel estimations of singleton users are almost perfect and, due to block fading channel assumption, the coefficients remain constant for all replicas. 
This assumption is realistic in all situations in which the slot time is short compared to the coherence time and in which the \ac{SC} strategy is applied. 
Making this assumption when no \ac{SC} protocol is enabled could instead be too optimistic.
\end{remark}

\section{Performance Evaluation}
\label{sec:NumericalResults}

In this section, we present numerical results about several \ac{PHY} layer processing strategies. 
To make fair comparisons in the context of \ac{mMTC} with reliability and latency constraints, we impose a common maximum latency and we plot the reliability in terms of \acl{PLR} $P_\mathrm{L}$ against the scalability represented by the number of simultaneously active user per frame $K_\mathrm{a}$. 
Moreover, we compare the techniques discussed in previous sections with some representative benchmarks, using also different \ac{MAC} protocols. 
In particular, we call ``baseline \ac{MAC}'' the standard repetition-based \ac{CSA} protocol with a constant number $r$ of replicas per packet transmitted in $r$ slots chosen uniformly at random in the frame. 
As a variation of this baseline protocol, we also adopt the recently proposed repetition-based \ac{CSA} with intra-frame \ac{SC} and \ac{ACK} messages \cite{ValChiPao:22}.


\subsection{Simulation Setup}
\label{subsec:SimSetUp}

We consider a system where users transmit payloads encoded with an $(n, k, t)$ narrow-sense binary \ac{BCH} code. 
A \ac{CRC} code is also used to validate decoded packets, avoiding that the \ac{SIC} procedure adds interference instead of subtracting it. 
Zero padding the \ac{BCH} codeword with a final bit, we can map the encoded bits onto a \ac{QPSK} constellation with Gray mapping, obtaining $N_\mathrm{D}$ symbols per codeword. 
The \ac{QPSK} symbol energy is normalized to one.
Simulations have been carried out with symbol rate $B_\mathrm{s} = 1$~Msps, $M = 256$ \ac{BS} antennas, $N_{\mathrm{P}}=64$ orthogonal pilot sequences, \ac{CSA} repetition degree $r=3$, and $\sigma_\mathrm{n}^2 = 0.1$.
We impose a maximum latency constraint $\Omega = 50$~ms. 
For a given maximum latency $\Omega$, the number of slots per frame $N_\mathrm{s}$ is equal to \cite{ValChiPao:22}
\begin{align}
\label{eq:NSlot}
    N_\mathrm{s} = 
    \left\lfloor \frac{\Omega \, B_\mathrm{s}}{2 \, (N_\mathrm{P} + N_\mathrm{D}) } \right\rfloor \, .
\end{align}
Note that the length of each orthogonal pilot equals the total number of available pilot sequences $N_{\mathrm{P}}$.
These sequences are constructed using Hadamard matrices.
Unless otherwise stated, we will consider that the coherence time is equal to the slot time.

\subsection{Numerical Results}
\label{subsec:NumRes}

\begin{figure}[t]
    \centering
    \includegraphics[width=0.99\columnwidth]{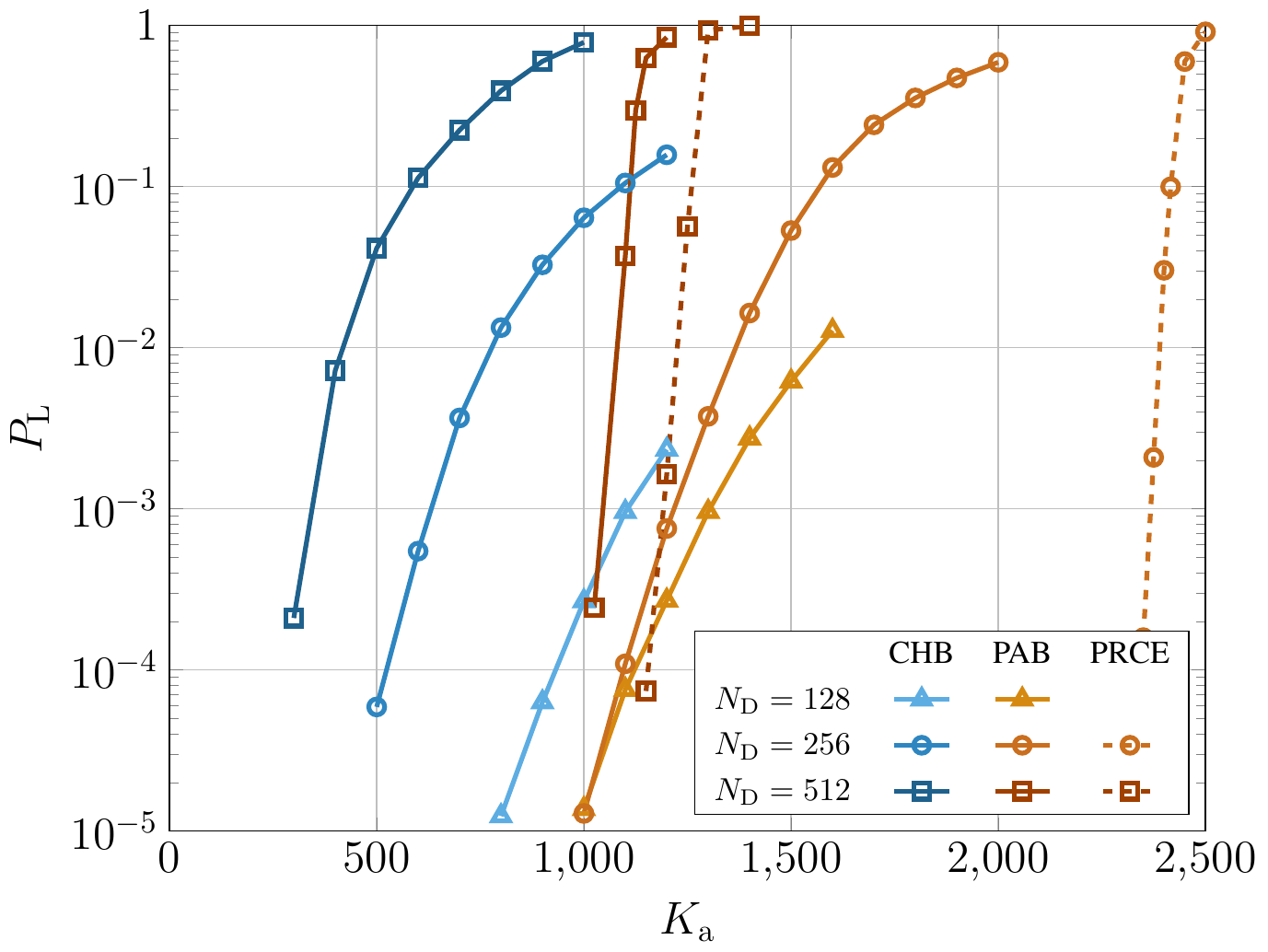}
    \caption{Packet loss rate values of schemes characterized by different \ac{SIC} techniques and payload sizes $N_\mathrm{D} = \{128, 256, 512 \}$. Baseline \ac{MAC} with $N_\mathrm{P} = 64$, $N_\mathrm{s} = \{130, 78, 43 \}$, and $M = 256$ antennas. Comparison between the \acs{CHB}, the proposed \acs{PAB} and the ideal \acs{SIC} case (\acs{PRCE}). For the sake of completeness, the \ac{PRCE} curve at $N_\mathrm{D} = 128$ intersect $P_\mathrm{L}^* = 10^{-3}$ around $K_\mathrm{a} = 4500$.}
    \label{fig:PLRVarND}
\end{figure}

In Fig.~\ref{fig:PLRVarND} we report the \ac{PLR} varying the symbol payload size $N_\mathrm{D}$ while keeping the rate of the \ac{BCH} code constant, for the \ac{CHB}, \ac{PAB}, and \ac{PRCE} (ideal) interference cancellation. 
To be precise, for $N_\mathrm{D} \in \{ 128, 256, 512 \}$ the corresponding \ac{BCH} codes are $(255, 207, 6)$, $(511, 421, 10)$, and $(1023, 843, 18)$. 
In this particular example, we adopt the baseline \ac{MAC} fixing $N_\mathrm{P} = 64$ leading to $N_\mathrm{s} \in \{130, 78, 43 \}$ in accordance with \eqref{eq:NSlot}.
As expected, the \ac{CHB} processing curves degrade when $N_\mathrm{D}$ increases due to the fact that the number of slots per frame $N_\mathrm{s}$ is decreasing. 
The same behavior can be observed for \ac{PRCE}. 
In the case of \ac{PAB} processing, instead, the trend is not so obvious. 
In fact, its performance tends to degrade when $N_\mathrm{s}$ decreases as for the other schemes, however, a gain in term of \ac{SIC} quality is also expected from \eqref{eq:VarHTilde}. 
In Fig.~\ref{fig:PLRVarND} we can see the gap between the \ac{PRCE} and the \ac{PAB} reduces, highlighting the effectiveness of the proposed technique in a complete scenario which accounts for both the \ac{PHY} and \ac{MAC} layers. 
In this particular example, these two effects counterbalance each other resulting in approximately $1000$ active users per frame at $P_\mathrm{L} = 10^{-4}$, for all $N_\mathrm{D}$ under examination using \ac{PAB}. 

\begin{figure}[t]
    \centering
    \includegraphics[width=0.99\columnwidth]{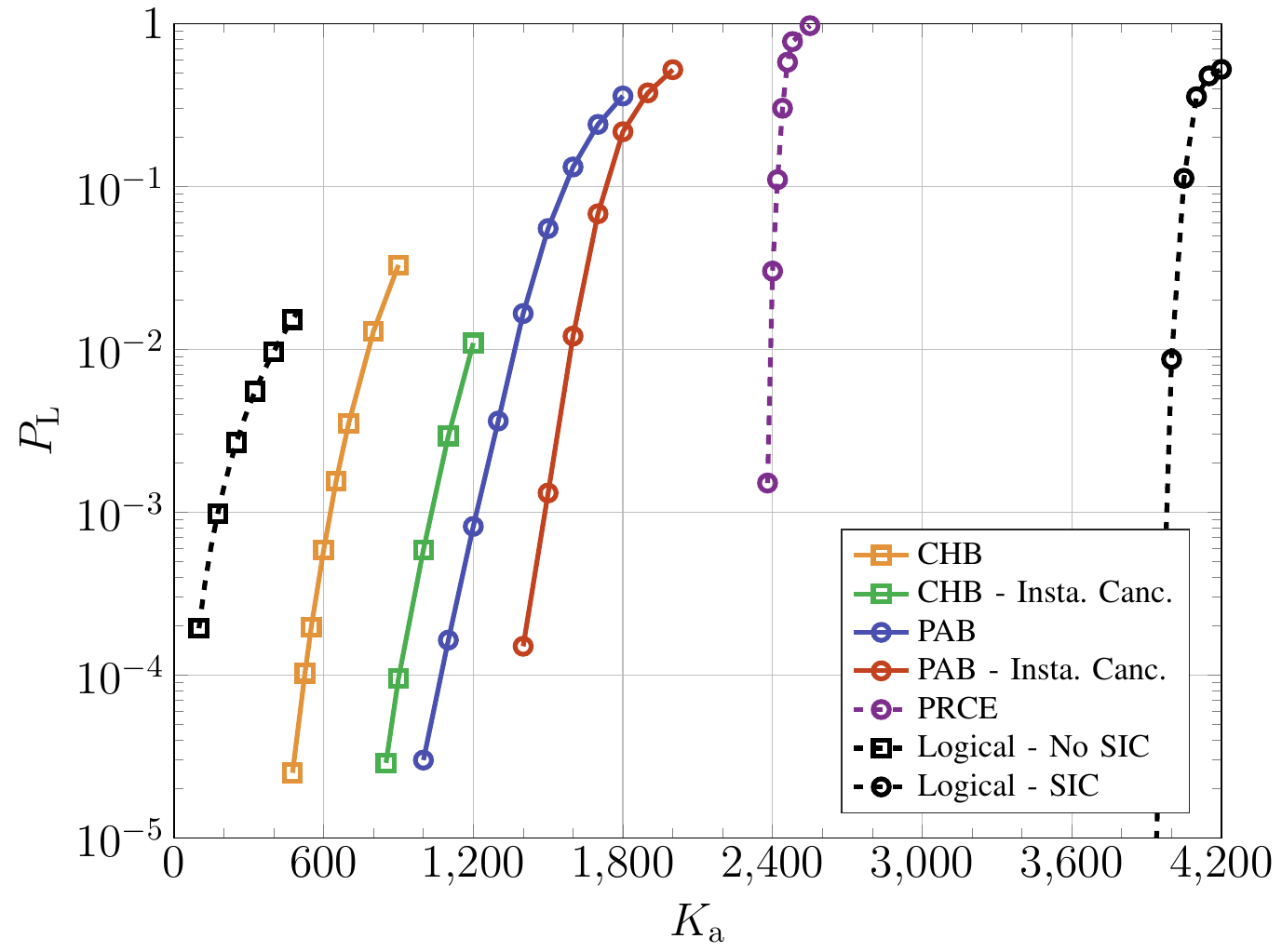}
    \caption{Packet loss rate comparison between different \ac{PHY} layer schemes, when a baseline \ac{MAC} protocol based on \ac{CSA} using repetition code with $r = 3$ is employed. Maximum latency $\Omega = 50$~ms, $M = 256$ antennas, $N_\mathrm{P} = 64$, $N_\mathrm{s} = 78$, and $N_\mathrm{D} = 256$.}
    \label{fig:PLR_baseline}
\end{figure}

In Fig.~\ref{fig:PLR_baseline} we plot a comparison between the \ac{CHB} and \ac{PAB} \ac{SIC} techniques, using the baseline \ac{MAC} protocol. 
We also apply instantaneous cancellation and plot the relative performance for both methods.
The number of payload symbols is set to $N_\mathrm{D} = 256$, leading to a $(511, 421, 10)$ \ac{BCH} code when an information payload of about $50$ Bytes is considered. The \ac{PAB} processing exhibits an improvement compared to the \ac{CHB}. 
This is motivated by the fact that \ac{PAB} subtractions have a beneficial effect on all users transmitting in a slot, while \ac{CHB} ones influence only the users employing a particular pilot.
Enabling instantaneous cancellation we obtain a remarkable performance boost in both \ac{SIC} algorithms. 
Targeting for example a \ac{PLR} $P_\mathrm{L} = 10^{-3}$, we see that the logical performance without \ac{SIC} achieves up to $180$ users per frame, the \ac{CHB} processing increases this number to $650$, and \ac{PAB} with instantaneous cancellation achieves a $K_\mathrm{a}$ of approximately $1500$. 
This $8\times$ increase in scalability motivates the interest on grant-free \ac{CRA} schemes under a realistic \ac{PHY} layer processing.

With reference to the same figure, we also point out the performance gap between a system performing realistic \ac{SIC} and two idealized schemes, the \ac{PRCE} and the logical one using \ac{SIC}. 
The \ac{PAB} and \ac{PRCE} curves rely on the same payload estimation, and for this reason their performance gap depends on channel estimation imperfections. 
At the same time, there is a remarkable gap between the \ac{PRCE} curve and the logical one using \ac{SIC} as a result of payload estimation non-idealities addressed in Section~\ref{subsec:DecSISNS}. 
Comparing the performance of actual schemes with these benchmarks reveals how neglecting the \ac{PHY} layer processing in real scenarios may lead to wrong conclusions and suboptimum optimizations.

\begin{figure}[t]
    \centering
    \includegraphics[width=0.99\columnwidth]{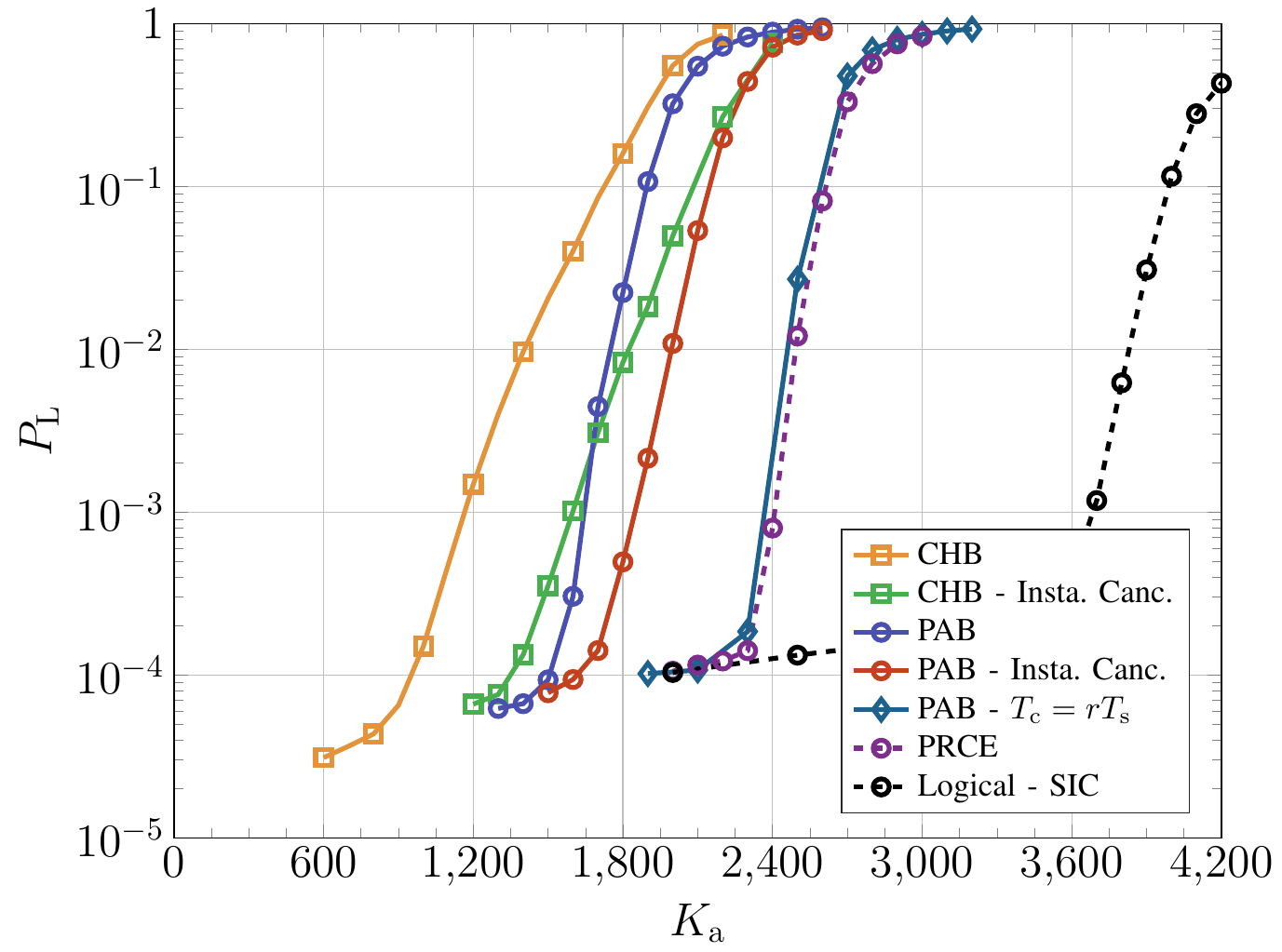}
    \caption{Packet loss rate comparison between different \ac{PHY} layer schemes, when intra-frame \acl{SC} and \acp{ACK} are enabled. \ac{CSA} using repetition code with $r = 3$ is employed, maximum latency $\Omega = 50$~ms, $M = 256$ antennas, $N_\mathrm{P} = 64$, $N_\mathrm{s} = 78$, and $N_\mathrm{D} = 256$.}
    \label{fig:PLR_ACKSC}
\end{figure}

In Fig.~\ref{fig:PLR_ACKSC} we report the performance of the same \ac{PHY} layer processing techniques of Fig.~\ref{fig:PLR_baseline}, when the \ac{MAC} access protocol recently presented in \cite{ValChiPao:22} is adopted. 
In particular, we consider intra-frame spatial coupling packet scheduling, where users are forced to transmit in adjacent slots. 
In addition, the \ac{BS} can send \ac{ACK} messages to notify successfully decoded users at the end of each slot to interrupt useless replica transmissions, resulting in interference attenuation and energy saving \cite{ValChiPao:22}. 
Despite the \ac{MAC} protocol change, the proposed \ac{PHY} layer processing techniques provide again a considerable performance improvement.

Let us now discuss how the \ac{PRCE} performance (i.e., same processing as \ac{PAB} but with ideal \ac{SIC}) can be approached using the proposed techniques. 
As anticipated when discussing Fig.~\ref{fig:PLRVarND}, one possibility to reduce the gap between \ac{PAB} and \ac{PRCE} is to increase $N_\mathrm{D}$. 
However, since we are considering a scenario where maximum latency is constrained, the degrading effect cause by $N_\mathrm{s}$ reduction is dominant. 
Hence, reaching \ac{PRCE} in this way could not give an overall boost in performance. 
Another case in which \ac{PRCE} curve can be reached is depicted in Fig.~\ref{fig:PLR_ACKSC}. 
So far we have considered block fading channel where the coherence time $T_\mathrm{c}$ is equal to the slot time $T_\mathrm{s}$. 
However, if the time slot is sufficiently small it is possible that, in some scenarios, the coherence time is several times $T_\mathrm{s}$.
Exploiting the characteristic of intra-frame spatial coupling, we can therefore have the same user channel coefficients among all the replicas ($T_\mathrm{c} \geq r \, T_\mathrm{s}$). 
Hence, when noise is sufficiently small, we can subtract interference of all replicas  using the channel estimates of singleton users, approaching ideal cancellation performance of \ac{PRCE}. 
Despite we are not using the payload information, we report this scheme as \ac{PAB} with $T_\mathrm{c} = r \, T_\mathrm{s}$ because it adopts iterative subtractions in \eqref{eq:PYmod}.

In Fig.~\ref{fig:PLR_baseline} and Fig.~\ref{fig:PLR_ACKSC} we remark the notable gap between \ac{PRCE} and the logical curve using \ac{SIC}. 
This gap is essentially due to the fact that singleton replicas (either the ones that arrived alone in a resource or those becoming singleton ones during the \ac{SIC} process) are not decoded with probability one and, thus, it is strictly related to Fig.~\ref{fig:InterfNS_Noise}. 
The analytical derivation developed in Section~\ref{subsec:DecSISNS}, and in particular the expression of $P_\mathrm{fail}$ in \eqref{eq:Pfail}, suggests possible solutions to narrow this gap: 
for example, we can increase the number of antennas $M$, or increase the error correction capability $t$ of the channel code (at the cost, however, of reducing the code rate and therefor the sum rate presented next). 
Some of these solutions are intuitively obvious, but the conducted analysis allows precisely quantifying the effect of a variation of each system parameter.
Another important factor which should be considered is the noise level. 
Nevertheless, since we have used $\sigma_n^2 = 0.1$ in the numerical evaluation, having a smaller noise level does not improve significantly the performance.

\begin{figure}[t]
    \centering
    \includegraphics[width=0.99\columnwidth]{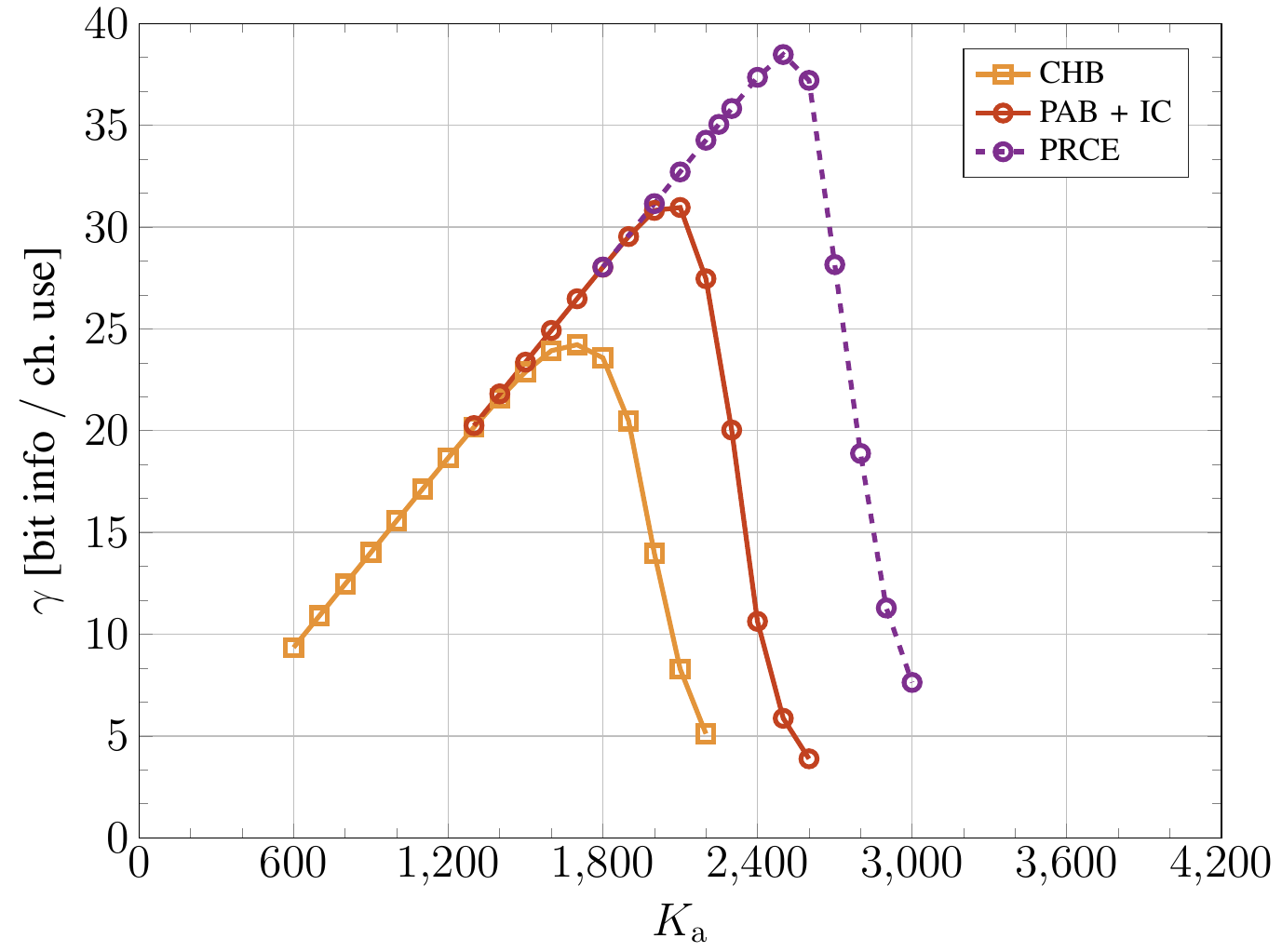}
    \caption{Sum rates in information bits per channel use of different \ac{PHY} layer schemes, when intra-frame \acl{SC} and \acp{ACK} are enabled. \ac{CSA} using repetition code with $r = 3$ is employed, maximum latency $\Omega = 50$~ms, $M = 256$ antennas, $N_\mathrm{P} = 64$, $N_\mathrm{D} = 256$, $N_\mathrm{s} = 78$, and $N_\mathrm{extra} = 33$.}
    \label{fig:sumrate}
\end{figure}

In Fig.~\ref{fig:sumrate} we show the sum rate in terms of information bits per channel use, defined as
\begin{align}
    \gamma = (1 - P_\mathrm{L})\, K_\mathrm{a}\, \frac{N_\mathrm{D} \, \log_2(\mathsf{M}) \, R_\mathrm{c} - N_\mathrm{extra}}{N_\mathrm{s} \, (N_\mathrm{P} + N_\mathrm{D})}
\end{align}
where $N_\mathrm{extra} = 33$, $R_\mathrm{c} = 421 / 511$, $\mathsf{M} = 4$, and other parameters are the same used in Fig.~\ref{fig:PLR_baseline} and Fig.~\ref{fig:PLR_ACKSC}.  
The parameter $N_\mathrm{extra}$ accounts for payload bits which are not used for information data as \ac{CRC} and zero padding bits.
In particular, we report the sum rates of some schemes using intra-frame spatial coupling packet scheduling with \acp{ACK}.
In this plot we observe that there exists an optimal $K_\mathrm{a}$ which maximizes the sum rate $\gamma$. 
However, the values of $K_\mathrm{a}$ yielding the largest $\gamma$ may correspond to values of reliability not fulfilling the requirements of next generation \ac{MMA} systems. 
On the other hand, the maximum value of the sum rate in information bits per second $\gamma_b = \gamma \, B_\mathrm{s}$ can be useful to design the backhaul communication network.

\section{Conclusions}\label{sec:conclusions}

Interference in grant-free access protocols poses a serious challenge for next generation \ac{MMA} systems. 
The use of \ac{CRA} with \acl{MPR} capabilities enabled by massive \ac{MIMO} and (randomly-chosen) orthogonal pilots improves system scalability, while allowing fulfillment of relatively-tightening latency and reliability constraints.
In this paper we showed how, for a given target reliability (i.e., \acl{PLR}), scalability is heavily reliant on the processing adopted at \ac{PHY} layer to perform interference subtraction.
The main conclusions of this paper can be summarized as: $i)$ the interference cancellation algorithm plays a very significant role in \ac{CRA}; $ii)$ it is important to efficiently schedule the subtraction operations due to cancellation imperfections; $iii)$ system design and analysis relying on collision-like channels may turn inaccurate.
For these reasons, we have proposed an interference cancellation algorithm and a scheduling strategy aiming at improving the overall performance.
For example, considering a target packet loss rate $P_\mathrm{L} = 10^{-3}$ and a requirement on maximum latency of $50$~ms, we found out that, employing both techniques, it is possible to achieve a $2.5\times$ scalability gain compared to the state-of-the-art and an $8\times$ gain compared to schemes without \ac{SIC}.

\section*{Acknowledgment}
This work has been carried out in the framework of the CNIT National Laboratory WiLab and the WiLab-Huawei Joint Innovation Center. 
The authors wish to thank Alberto Faedi for his work on software simulator implementation.


\appendices
\section{Interference Analysis on General Modulation and Coding Schemes} \label{app:GeneralizationModCod}
To generalize the approach to arbitrary modulation and coding schemes, we observe that \eqref{eq:EqChModel}, for a given $\| \V{h}_\ell \|^2$, defines an additive Gaussian channel, with a ratio between the average energy per symbol and the one-sided noise power spectral density given by
\begin{align}
\label{eq:EsN0}
    \frac{E_\mathrm{s}}{N_0} = \frac{\| \V{h}_\ell \|^4}{ \Var{\tilde{I}_j}} = \frac{w^2}{ 4\Var{\tilde{I}_j}}\,.
\end{align}
Thus, it is possible to replace \eqref{eq:PfailGivenw} by substituting $P_{\mathrm{fail}|w}$ with the relation between the codeword error probability and $E_\mathrm{s}/{N_0}$ for the modulation and coding scheme of interest. For example, we can use the error probability vs. $E_\mathrm{s}/{N_0}$ derived for \ac{LDPC} or Turbo codes, with \ac{QPSK}  modulation. This generalization can be useful to construct analytical designing tools for \ac{CRA} schemes in realistic scenario, as done in \cite{Valentini2022:Joint}. 
Note that, if some activity detection algorithm is employed, the decoder will work with the knowledge of the signal-to-noise ratio $E_\mathrm{s}/{N_0}$, as this is related to the actual number of active users, $|\mathcal{A}|$. Otherwise, the decoder should be designed to work sub-optimally, with an unknown signal-to-noise ratio.   
In most cases the codeword error probability vs. $E_\mathrm{s}/{N_0}$ function cannot be found analytically, and Monte Carlo simulation should be used. 

\section{Analytical Performance without SIC} \label{sec:WithoutSIC}
In this appendix we derive the average number of successfully decoded users, assuming a collision channel over resources model, when no \ac{SIC} is performed. 
This analysis can be used as a benchmark to evaluate the effectiveness of the proposed \ac{SIC} strategy. 
To keep a clean and compact notation, we denote the probability that a random variable $\rv{A}$ takes the value $a$, $\Prob{\rv{A} = a}$, as $P(a)$. 
Similarly, we write $P(a, b\,|\,c)$ to indicate the probability $\Prob{\rv{A} = a, \rv{B} = b \,|\, \rv{C} = c}$, and $\Prob{\mathcal{E}}$ to indicate the probability that an event $\mathcal{E}$ holds.

Let us consider the following problem. 
There are $K_\mathrm{a}$ active devices, each of which transmits $r$ replicas of its packet into a frame composed of $N_\mathrm{s}$ slots. 
The device can put no more than one replica in each slot, and in each slot it can choose between $N_\mathrm{P}$ possible orthogonal pilots. 
Therefore we can describe the frame as a grid of $R = N_\mathrm{s} \cdot N_\mathrm{P}$ resources. 
Defining as \emph{uncollided} a user, any replica of which has arrived alone in a resource, under a collision channel model the number of successful users in the current frame equals the number of uncollided ones.
We can write the total number of uncollided users as
\begin{align}
    \rv{X} = \rv{X}_1 + \rv{X}_2 + \dots + \rv{X}_{K_\mathrm{a}}
\end{align}
where
\begin{align}
    \rv{X}_i = \begin{dcases}1 & \text{if at least one replica of user $i$ is uncollided}\\
    0 & \text{otherwise}\,.
    \end{dcases}
\end{align}
The average number of uncollided users can therefore be written as
\begin{align}
    \E{\rv{X}} = \sum_{i = 0}^{K_\mathrm{a}} \E{\rv{X}_i} = K_\mathrm{a} \cdot \Prob{\rv{X}_i = 1}\,.
\end{align}
Denoting by $\mathcal{U}$ the event that the generic replica transmitted by an active user arrives alone in a resource, we have
\begin{align}
    \Prob{\rv{X}_i = 1} &= 1 - (1 - \Prob{\mathcal{U}})^r\,.
\end{align}

Next, let us focus on a single replica from an active device. 
Let the considered replica be interfered by $\rv{J}$ replicas transmitted by other devices that have chosen the same slot.
By law of total probability we can write
\begin{align}
    \Prob{\mathcal{U}} &= \sum_j \Prob{\mathcal{U}, j}  = \sum_j \Prob{\mathcal{U}| j} \, P(j)
\end{align}
where it is immediate to see that 
\begin{align}
    \Prob{\mathcal{U}|j} = \left(\frac{N_\mathrm{P} - 1}{N_\mathrm{P}}\right)^j\,.
\end{align}
To derive $P(j)$, we firstly write the probability that none of the $r$ replicas is transmitted in a specific slot as
\begin{align}
    \frac{(N_\mathrm{s}-1)\dots(N_\mathrm{s}-r)}{N_\mathrm{s} \dots (N_\mathrm{s}-r-1)} = 1- \frac{r}{N_\mathrm{s}}\,.
\end{align}
Consequentially, we can derive $P(j)$ as
\begin{align}
    P(j) = \binom{K_\mathrm{a}-1}{j} \left( \frac{r}{N_\mathrm{s}} \right)^j \left( 1 - \frac{r}{N_\mathrm{s}} \right)^{K_\mathrm{a}-1-j}
\end{align}
and conclude that
\begin{align}
	\Prob{\mathcal{U}} &= \sum_{j = 0}^{K_\mathrm{a}-1} \binom{K_\mathrm{a}-1}{j} \left( \frac{r}{N_\mathrm{s}} \, \frac{N_\mathrm{P} - 1}{N_\mathrm{P}} \right)^j \left( 1 - \frac{r}{N_\mathrm{s}} \right)^{K_\mathrm{a}-1-j} \nonumber \\
	&= \left(1 - \frac{r}{N_\mathrm{s}N_\mathrm{P}}\right)^{K_\mathrm{a}-1}	\,.
\end{align}
Finally, in absence of \ac{SIC} the packet loss probability is
\begin{align}
    P_\mathrm{L, noSIC} &= 1 - \frac{\E{X}}{K_\mathrm{a}} = \left(1 - \left(1 - \frac{r}{N_\mathrm{s}N_\mathrm{P}}\right)^{K_\mathrm{a}-1}\right)^r\,.
\end{align}

\bibliographystyle{IEEEtran}
\bibliography{Files/IEEEabrv,Files/StringDefinitions,Files/StringDefinitions2,Files/refs}

\begin{thebibliography}{10}
\providecommand{\url}[1]{#1}
\csname url@samestyle\endcsname
\providecommand{\newblock}{\relax}
\providecommand{\bibinfo}[2]{#2}
\providecommand{\BIBentrySTDinterwordspacing}{\spaceskip=0pt\relax}
\providecommand{\BIBentryALTinterwordstretchfactor}{4}
\providecommand{\BIBentryALTinterwordspacing}{\spaceskip=\fontdimen2\font plus
\BIBentryALTinterwordstretchfactor\fontdimen3\font minus
  \fontdimen4\font\relax}
\providecommand{\BIBforeignlanguage}[2]{{%
\expandafter\ifx\csname l@#1\endcsname\relax
\typeout{** WARNING: IEEEtran.bst: No hyphenation pattern has been}%
\typeout{** loaded for the language `#1'. Using the pattern for}%
\typeout{** the default language instead.}%
\else
\language=\csname l@#1\endcsname
\fi
#2}}
\providecommand{\BIBdecl}{\relax}
\BIBdecl

\bibitem{Sachs2016:Machine}
J.~Sachs, P.~Popovski, A.~Höglund, D.~Gozalvez-Serrano, P.~Fertl, M.~Dohler,
  and T.~Nakamura, ``Machine-type communications,'' in \emph{5G Mobile and
  Wireless Communications Technology}, A.~Osseiran, J.~F. Monserrat, and
  P.~Marsch, Eds.\hskip 1em plus 0.5em minus 0.4em\relax Cambridge University
  Press, 2016, ch.~4, p. 77–106.

\bibitem{Shariatmadari2015:Machine}
H.~Shariatmadari, R.~Ratasuk, S.~Iraji, A.~Laya, T.~Taleb, R.~J{\:a}ntti, and
  A.~Ghosh, ``Machine-type communications: {C}urrent status and future
  perspectives toward {5G} systems,'' \emph{IEEE Commun. Mag.}, vol.~53, no.~9,
  pp. 10--17, Sep. 2015.

\bibitem{Bockelmann2016:Massive}
C.~Bockelmann, N.~Pratas, H.~Nikopour, K.~Au, T.~Svensson, C.~Stefanovic,
  P.~Popovski, and A.~Dekorsy, ``Massive machine-type communications in {5G}:
  {P}hysical and {MAC}-layer solutions,'' \emph{IEEE Commun. Mag.}, vol.~54,
  no.~9, pp. 59--65, Sep. 2016.

\bibitem{Lien2011:Toward}
S.-Y. Lien, K.-C. Chen, and Y.~Lin, ``Toward ubiquitous massive accesses in
  {3GPP} machine-to-machine communications,'' \emph{IEEE Commun. Mag.},
  vol.~49, no.~4, pp. 66--74, Apr. 2011.

\bibitem{Wu2020:Massive}
Y.~Wu, X.~Gao, S.~Zhou, W.~Yang, Y.~Polyanskiy, and G.~Caire, ``Massive access
  for future wireless communication systems,'' \emph{IEEE Wireless Commun.},
  vol.~27, no.~4, pp. 148--156, Aug. 2020.

\bibitem{wolf1981:coding}
J.~Wolf, ``Coding techniques for multiple access communication channels,'' in
  \emph{New Concepts in Multi-User Communication}, J.~Skwirzynski, Ed.\hskip
  1em plus 0.5em minus 0.4em\relax Alphen an de Rijn, The Netherlands: Sijthoff
  \& Noordhoff, 1981, pp. 83--103.

\bibitem{Gallager1985:Perspective}
R.~Gallager, ``A perspective on multiaccess channels,'' \emph{Proc. IEEE},
  vol.~31, no.~2, pp. 124--142, Mar. 1985.

\bibitem{Mathys1990:class}
P.~Mathys, ``A class of codes for a {$T$} active users out of {$N$}
  multiple-access communication system,'' \emph{{IEEE} Trans. Inf. Theory},
  vol.~36, no.~6, pp. 1206--1219, Nov. 1990.

\bibitem{Durisi2016:Toward}
G.~Durisi, T.~Koch, and P.~Popovski, ``Toward massive, ultrareliable, and
  low-latency wireless communication with short packets,'' \emph{Proc. IEEE},
  vol. 104, no.~9, pp. 1711--1726, Aug. 2016.

\bibitem{Polyanskiy2017:Perspective}
Y.~Polyanskiy, ``A perspective on massive random-access,'' in \emph{2017 IEEE
  Int. Symp. Inf. Theory}, Aachen, Germany, Jun. 2017, pp. 2523--2527.

\bibitem{Chen2017:Capacity}
X.~Chen, T.-Y. Chen, and D.~Guo, ``Capacity of {G}aussian many-access
  channels,'' \emph{IEEE Trans. Inf. Theory}, vol.~63, no.~6, pp. 3516--3539,
  Jun. 2017.

\bibitem{Ngo2021:Random_user_activity}
K.-H. Ngo, A.~Lancho, G.~Durisi, and A.~Graell~i Amat, ``Massive uncoordinated
  access with random user activity,'' in \emph{Proc. 2021 IEEE Int. Symp. Inf.
  Theory}, Melbourne, Australia, Jun. 2021.

\bibitem{Paolini22:Irregular}
E.~Paolini, L.~Valentini, V.~Tralli, and M.~Chiani, ``Irregular repetition
  slotted {ALOHA} in an information-theoretic setting,'' in \emph{Proc. 2022
  IEEE Int. Symp. Inf. Theory}, Espoo, Finland, Jun. 2022.

\bibitem{Hasan2013:random}
M.~Hasan, E.~Hossain, and D.~Niyato, ``Random access for machine-to-machine
  communication in {LTE}-advanced networks: Issues and approaches,''
  \emph{{IEEE} Commun. Mag.}, vol.~51, no.~6, pp. 86--93, Jun. 2013.

\bibitem{Liu2018:sparse}
L.~Liu, E.~G. Larsson, W.~Yu, P.~Popovski, C.~Stefanovic, and E.~De~Carvalho,
  ``Sparse signal processing for grant-free massive connectivity: A future
  paradigm for random access protocols in the internet of things,''
  \emph{{IEEE} Signal Process. Mag.}, vol.~35, no.~5, pp. 88--99, Sep. 2018.

\bibitem{Chen2020:massive}
X.~Chen, D.~W.~K. Ng, W.~Yu, E.~G. Larsson, N.~Al-Dhahir, and R.~Schober,
  ``Massive access for {5G} and beyond,'' \emph{{IEEE} J. Sel. Areas Commun.},
  vol.~39, no.~3, pp. 615--637, Mar. 2021.

\bibitem{Gui2020:6G}
G.~Gui, M.~Liu, F.~Tang, N.~Kato, and F.~Adachi, ``6{G}: {O}pening new horizons
  for integration of comfort, security, and intelligence,'' \emph{{IEEE}
  Wireless Commun.}, vol.~27, no.~5, pp. 126--132, Oct. 2020.

\bibitem{Kalalas2020:massive}
C.~Kalalas and J.~Alonso-Zarate, ``Massive connectivity in {5G} and beyond:
  {T}echnical enablers for the energy and automotive verticals,'' in
  \emph{Proc. 2020 2nd 6G Wireless Summit}, Levi, Finland, Mar. 2020.

\bibitem{Pokhrel2020:Towards}
S.~R. Pokhrel, J.~Ding, J.~Park, O.-S. Park, and J.~Choi, ``Towards enabling
  critical {mMTC}: {A} review of {URLLC} within {mMTC},'' \emph{{IEEE} Access},
  vol.~8, pp. 131\,796--131\,813, Jul. 2020.

\bibitem{Liu2018:massivePt1}
L.~Liu and W.~Yu, ``Massive connectivity with massive {MIMO}—part {I}: Device
  activity detection and channel estimation,'' \emph{{IEEE} Trans. Signal
  Process.}, vol.~66, no.~11, pp. 2933--2946, Mar. 2018.

\bibitem{Sor2018:coded}
J.~H. S{\o}rensen, E.~De~Carvalho, {\v{C}}.~Stefanovic, and P.~Popovski,
  ``Coded pilot random access for massive {MIMO} systems,'' \emph{{IEEE} Trans.
  Wireless Commun.}, vol.~17, no.~12, pp. 8035--8046, Dec. 2018.

\bibitem{Fengler2019:grant-free}
A.~Fengler, S.~Haghighatshoar, P.~Jung, and G.~Caire, ``Grant-free massive
  random access with a massive {MIMO} receiver,'' in \emph{2019 53rd Asilomar
  Conf. Signals, Systems, Computers}, Pacific Grove, CA, USA, Nov. 2019, pp.
  23--30.

\bibitem{Han2020:grant-free}
H.~Han, Y.~Li, W.~Zhai, and L.~Qian, ``A grant-free random access scheme for
  {M2M} communication in massive {MIMO} systems,'' \emph{{IEEE} Internet Things
  J.}, vol.~7, no.~4, pp. 3602--3613, Apr. 2020.

\bibitem{Abebe2021:MIMO}
A.~T. Abebe and C.~G. Kang, ``{MIMO}-based reliable grant-free massive access
  with {QoS} differentiation for 5{G} and beyond,'' \emph{{IEEE} J. Sel. Areas
  Commun.}, vol.~39, no.~3, pp. 773--787, Mar. 2021.

\bibitem{Choi2022:grant}
J.~Choi, J.~Ding, N.-P. Le, and Z.~Ding, ``Grant-free random access in
  machine-type communication: Approaches and challenges,'' \emph{{IEEE}
  Wireless Commun.}, vol.~29, no.~1, pp. 151--158, Feb. 2022.

\bibitem{Decurninge2021:Tensor}
A.~Decurninge, I.~Land, and M.~Guillaud, ``Tensor-based modulation for
  unsourced massive random access,'' \emph{{IEEE} Wireless Commun. Lett.},
  vol.~10, no.~3, pp. 552--556, Mar. 2021.

\bibitem{Liu22:unsourced}
J.~Liu and X.~Wang, ``Unsourced multiple access based on sparse tanner
  graph-efficient decoding, analysis, and optimization,'' \emph{{IEEE} J. Sel.
  Areas Commun.}, vol.~40, no.~5, pp. 1509--1521, May 2022.

\bibitem{casini2007:contention}
E.~Casini, R.~{De Gaudenzi}, and O.~{del Rio Herrero}, ``Contention resolution
  diversity slotted {ALOHA} ({CRDSA}): {A}n enhanced random access scheme for
  satellite access packet networks,'' \emph{{IEEE} Trans. Wireless Commun.},
  vol.~6, no.~4, pp. 1408--1419, Apr. 2007.

\bibitem{liva2011:irsa}
G.~Liva, ``Graph-based analysis and optimization of contention resolution
  diversity slotted {ALOHA},'' \emph{{IEEE} Trans. Commun.}, vol.~59, no.~2,
  pp. 477--487, Feb. 2011.

\bibitem{paolini2015:csa}
E.~Paolini, G.~Liva, and M.~Chiani, ``Coded slotted {ALOHA}: A graph-based
  method for uncoordinated multiple access,'' \emph{{IEEE} Trans. Inf. Theory},
  vol.~61, no.~12, pp. 6815--6832, Dec. 2015.

\bibitem{paolini2015:magazine}
E.~Paolini, {\v C}.~Stefanovi{\'c}, G.~Liva, and P.~Popovski, ``Coded random
  access: {A}pplying codes on graphs to design random access protocols,''
  \emph{{IEEE} Commun. Mag.}, vol.~53, no.~6, pp. 144--150, Jun. 2015.

\bibitem{clazzer2018:combining}
F.~Clazzer, C.~Kissling, and M.~Marchese, ``Enhancing contention resolution
  {ALOHA} using combining techniques,'' \emph{{IEEE} Trans. Commun.}, vol.~66,
  no.~6, pp. 2576--2587, Jun. 2018.

\bibitem{Berioli2016:Modern}
M.~Berioli, G.~Cocco, G.~Liva, and A.~Munari, ``Modern random access
  protocols,'' \emph{Foundations and Trends in Networking}, vol.~10, no.~4, pp.
  317--446, 2016.

\bibitem{Munari2021:age}
A.~Munari, ``Modern random access: {A}n age of information perspective on
  irregular repetition slotted {ALOHA},'' \emph{{IEEE} Trans. Commun.},
  vol.~69, no.~6, pp. 3572--3585, Jun. 2021.

\bibitem{ValChiPao:22}
L.~Valentini, M.~Chiani, and E.~Paolini, ``Massive grant-free access with
  massive {MIMO} and spatially coupled replicas,'' \emph{{IEEE} Trans.
  Commun.}, vol.~70, no.~11, pp. 7337--7350, 2022.

\bibitem{Mahmood2020:Six}
N.~H. Mahmood, H.~Alves, O.~A. López, M.~Shehab, D.~P.~M. Osorio, and
  M.~Latva-Aho, ``Six key features of machine type communication in 6{G},'' in
  \emph{Proc. 2020 2nd 6G Wireless Summit}, Levi, Finland, Mar. 2020.

\bibitem{Gha2013:irregular}
M.~Ghanbarinejad and C.~Schlegel, ``Irregular repetition slotted {ALOHA} with
  multiuser detection,'' in \emph{Proc. 2013 10th Annual Conf. Wireless
  On-demand Netw. Systems Services}, Banff, AB, Canada, Mar. 2013.

\bibitem{stefanovic2018:multipacket}
{\v C}.~Stefanovi{\' c}, E.~Paolini, and G.~Liva, ``Asymptotic performance of
  coded slotted {ALOHA} with multipacket reception,'' \emph{{IEEE} Commun.
  Lett.}, vol.~22, no.~1, pp. 105--108, Jan. 2018.

\bibitem{Valentini2022:Joint}
L.~Valentini, M.~Chiani, and E.~Paolini, ``A joint {PHY} and {MAC} layer design
  for coded random access with massive {MIMO},'' in \emph{Proc. 2022 IEEE
  Global Commun. Conf.}, Rio di Janeiro, Brazil, Dec. 2022.

\bibitem{Valentini2022:Impact}
L.~Valentini, A.~Faedi, M.~Chiani, and E.~Paolini, ``Impact of interference
  subtraction on grant-free multiple access with massive {MIMO},'' in
  \emph{Proc. 2022 IEEE Int. Conf. Commun.}, Seoul, South Korea, May 2022.

\bibitem{Bjo2017:MIMObook}
E.~Bj{\"o}rnson, J.~Hoydis, L.~Sanguinetti \emph{et~al.}, ``Massive {MIMO}
  networks: Spectral, energy, and hardware efficiency,'' \emph{Foundations and
  Trends{\textregistered} in Signal Processing}, vol.~11, no. 3-4, pp.
  154--655, 2017.

\bibitem{Con2005:MQAM}
A.~Conti, M.~Win, and M.~Chiani, ``Invertible bounds for {M-QAM} in {Rayleigh}
  fading,'' \emph{{IEEE} Trans. Wireless Commun.}, vol.~4, no.~5, pp.
  1994--2000, Sep. 2005.

\end{thebibliography}

\end{document}